\documentclass[useAMS,usenatbib,usegraphicx]{mn2e}
\usepackage{times}
\usepackage{subfigure}
\usepackage{float}
\usepackage{graphicx}
\usepackage{pdfpages}
\usepackage{adjustbox}
\usepackage{amsmath}
\usepackage{amssymb}
\usepackage[utf8]{inputenc}
\usepackage{newunicodechar}
\usepackage{array}
\usepackage{multirow}
\usepackage{pdflscape}

\newcommand{\ms}{M$_{\odot}$}
\newcommand{\rs}{R$_{\odot}$}

\newunicodechar{−}{-}

\newcolumntype{+}{>{\global\let\currentrowstyle\relax}}
\newcolumntype{^}{>{\currentrowstyle}}

\title[Common envelope simulations and observations]
{Speaking with one voice: simulations and observations discuss the common envelope $\alpha$ parameter}
\author[Iaconi \& De Marco]{Roberto Iaconi $^{1,2,3}$\thanks{E-mail: roberto.iaconi@kusastro.kyoto-u.ac.jp} \thanks{JSPS International Research Fellow (Graduate School of Science, Kyoto University)} and Orsola De Marco $^{2,3}$\\
$^1$Department of Astronomy, Kyoto University, Kitashirakawa-Oiwake-cho, Sakyo-ku, Kyoto 606-8502, Japan 0000-0002-1940-1950\\
$^{2}$Department of Physics \& Astronomy, Macquarie University, Sydney, NSW 2109, Australia\\
$^{3}$Astronomy, Astrophysics and Astrophotonics Research Centre, Macquarie University, Sydney, NSW 2109, Australia\\}

\begin{document}

\date{\today}

\pagerange{\pageref{firstpage}--\pageref{lastpage}} \pubyear{2018}

\maketitle

\label{firstpage}

\begin{abstract}
We present a comparative study between the results of most hydrodynamic simulations of the common envelope binary interaction to date and observations of post common envelope binaries. The goal is to evaluate whether this dataset indicates the existence of a formula that may predict final separations of post-common envelope systems as a function of pre-common envelope parameters. Some of our conclusions are not surprising while others are more subtle. We find that: (i) Values of the final orbital separation derived from common envelope simulations must at this time be considered upper limits. Simulations that include recombination energy do not seem to have systematically different final separations; these and other simulations imply $\alpha_{\rm CE} < 0.6-1.0$. At least one simulation, {applicable to double-degenerate systems}, implies $\alpha_{\rm CE} < 0.2$. (ii) Despite large reconstruction errors, the post-RGB observations reconstructed parameters are in agreement with some of the simulations. The post-AGB observations behave instead as if they had a systematically lower value of $\alpha_{\rm CE}$. The lack of common envelope simulations with low mass AGB stars leaves us with no insight as to why this is the case.  (iii) The smallest mass companion that survives the common envelope with intermediate mass giants is 0.05-0.1~\ms. (iv) Observations of binaries with separations larger than $\sim$10~\rs, tend to have high $M_2/M_1$ mass ratios and may go through a relatively long phase of unstable Roche lobe mass transfer followed by a weakened common envelope (or with no common envelope at all). (v) The effect of the spatial resolution and of the softening length on simulation results remains poorly quantified.
\end{abstract}

\begin{keywords}
stars: AGB and post-AGB - stars: evolution - binaries: close - hydrodynamics - methods: numerical, analytical
\end{keywords}


\section{Introduction}
\label{sec:intro}
The common envelope interaction (hereafter CE; \citealt{Paczynski1976}, \citealt{Ivanova2013}) between an expanding giant and a close companion is likely at the heart of a range of binary classes and phenomena, including type Ia supernovae and the emission of gravitational waves. Yet a description of this phenomenon that makes testable predictions has eluded us so far.
CE simulations have intensified in recent years with more codes being applied to this problem \citep[e.g.,][to cite a representative few]{Rasio1996,Sandquist1998,Passy2012,Nandez2015,Ohlmann2016,Chamandy2018}. However, the complexity of the simulations and the long compute times have limited the number of simulated cases and have effectively precluded convergence tests.

One of the main hurdles facing simulations is that if the simulation includes solely gravity and an ideal gas equation of state, the in-spiral unbinds less than half the envelope mass early on, but the entire envelope is lifted significantly and, in some cases, brought into corotation, such that the reduced gravitational drag induces a slowing down of the in-spiral \citep{Ricker2008,Staff2016a,Iaconi2018}. The inclusion of recombination energy in simulations, results in the full unbinding of the envelope for lower mass stars \citep{Nandez2016}, but the availability of the entire recombination energy budget to do work has been questioned by \citet[][see also \citealt{Ivanova2018} and \citealt{Grichener2018}]{Soker2018}. Even allowing the entire recombination energy budget to do work, there are still problems in ejecting the envelope for stars more massive than $\sim$2~\ms \ (\citealt{Iaconi2018}). The inclusion of additional physics in the simulations, such as jets (\citealt{Shiber2017}, \citealt{shiber2018}, \citealt{Shiber2019}) or convection (\citealt{wilson2019}) may complicate the issue further. 

The ``$\alpha$-formalism" \citep[e.g.,][]{Webbink1984} implies that the final separation of the post-CE binary is related to stellar and binary parameters but mediated by an efficiency factor, often called $\alpha_{\rm CE}$ in the literature, that measures the fraction of released orbital energy that is used to do work and unbind the envelope gas:
\begin{equation}
E_{\rm bin} = - \alpha_{\rm CE} G \left( \frac{M_{\rm c}M_2}{2 a_{\rm f}} - \frac{M_1 M_2}{2 a_{\rm i}} \right) \ ,
\label{eq:alpha_formalism}
\end{equation}
where $E_{\rm bin}$ is the envelope binding energy, $G$ is the gravitational constant,  $M_{\rm c}$, $M_1$ and $M_2$ are the primary's core mass, the primary's total mass at the time of the CE interaction and the companion's mass, respectively; $a_{\rm i}$ and $a_{\rm f}$ are the initial and final orbital separation, respectively. Using simulations to derive how $\alpha_{\rm CE}$ depends on the systems' parameters has proven difficult, primarily because of the lack of envelope unbinding renders the simulated final separations untrustworthy (Equation~\ref{eq:alpha_formalism} is technically not useful {\it unless} the envelope is fully unbound), and second because simulations are time consuming and do not cover sufficient parameter space. However, if we could determine the reliability of the final separations in simulations, we may yet find a way in which populations synthesis simulations can directly predict final separations of any kind of binary that goes through the CE interaction, as a function of stellar and binary parameters, bypassing the $\alpha$-formalism altogether.  

Several authors have instead quantified $\alpha_{\rm CE}$ by using observations of post-CE binary systems, alongside a reconstruction method that allows to determine the stellar structure at the time of the interaction. The error bars are substantial and some seemingly discrepant results have generated some confusion: $\alpha_{\rm CE}$  could be constant \citep[e.g.,][]{Zorotovic2010}, or a function of parameters such as the mass ratio \citep{DeMarco2011}. Finally, by using a population synthesis method, $\alpha_{\rm CE}$ has been suggested to have a dependency on parameters \citep{Politano2007} and, by using 1D stellar models with convection, $\alpha_{\rm CE}$ has been determined to have a far more complex and less predictable dependence on parameters  \citep[][who reconciled the results of \citealt{Zorotovic2011} and of \citealt{DeMarco2011}]{wilson2019}.

In this work we bring {\it almost all} 3D hydrodynamic simulations {\it and} observations of single degenerate, post-CE binaries to bear on this topic\footnote{We have decided not to attempt a reconstruction of double degenerate systems, which may have suffered more than one mass transfer episode or even two common envelope phases.}. We do so to validate theoretical studies as well as to determine whether we can find a signal in the combination of all data points. An earlier comparison was performed by \citet{Sandquist2000}, who carried out 5 CE simulations with the specific goal of comparing the simulations with 8 observed systems, 4 double degenerate and 4 single degenerates. With a different approach, \citet{Nandez2016} compared their simulations with a set of sdOB and post-AGB observations. Here we bring the corpus of simulations carried out to date to bear on a large sample of observed systems. This can be considered an extension of the work started by \citet{Iaconi2017} where all simulations available at that time were compared, but where the emphasis was not on the observations. 

This paper is structured as follows. In Section~\ref{sec:observations} we discuss the observational samples, the processing of the data and their characteristics. In Section~\ref{sec:simulations} we do the same for the 3D hydrodynamic simulations in the literature. In Section~\ref{sec:obs_sim_comparison} we compare observations and simulations, highlight similarities and differences of the two populations and discuss possible problems related to the comparison. Finally, in Section~\ref{sec:conclusions}, we summarise and conclude. 
In Appendix~\ref{sec:observations_data} we list the observational data and the quantities derived from them, together with similar data for simulations, where available.


\section{Post-common envelope observations}
\label{sec:observations}
In this section we consider observations of post-CE binaries from the literature. We adopt the sample published by \citet{Zorotovic2011b} in its entirety. We also study a second sample that is a mixture of observations from various publications. Below we discuss the characteristics of these data, our calculations to reconstruct binary parameters at the time of the CE interaction and give details of some individual objects.

\subsection{The observational and reconstructed data}
\label{ssec:obs_data}
The sample of \citet{Zorotovic2011b} has been considered in its entirety. This sample has considerable overlap with the sample of \citet{Zorotovic2010}, that was previously considered by \citet{Passy2012} and \citet{Iaconi2017}. We report the data from \citet{Zorotovic2011b} in Table~\ref{tab:obs_zorotovic}, where, in addition to the name of the objects we list the mass ratio of the binary at the time of the CE onset ($q$= $M_2$/$M_1$), the mass of the primary at the moment of CE ($M_1$), the observed mass of the WD ($M_{\mathrm{c}}$, where we list the error when its size makes the RGB or AGB interpretation uncertain), the radius of the primary at the moment of CE ($R_1$), the observed companion mass ($M_2$; note that using the observed value of $M_2$ assumes that the companion accretes an insignificant amount of mass during the CE interaction), the main sequence companion radius ($R_2$), the ratio between $R_2$ and the Roche lobe radius of the companion for the observed system ($R_2/R_{\mathrm{RL,2}}$), the observed final separation ($a_{\mathrm{f}}$), the absolute value of the binding energy of the primary at the moment of CE ($|E_{\mathrm{bin}}|$) and the change in orbital energy (where for the initial separation, $a_{\rm i}$ we have adopted the separation at the time of the onset of Roche lobe overflow).

The second data set we use is far less homogeneous than the previous one and includes in part that used by \citet[][except for HD149382, which was never confirmed as a binary, \citealt{Norris2011}]{DeMarco2011}, with the addition of other relevant objects (Table~\ref{tab:obs_demarco}, where the columns are the same as those of Table~\ref{tab:obs_zorotovic}, except for the last column that gives the original reference for each object). In this second data set we only list objects not already present in the list of \citet[][Table~\ref{tab:obs_zorotovic}]{Zorotovic2011b}. We also include central stars of planetary nebulae (PN). For these the CE must have happened in the recent past because the nebula is usually younger than 10,000 years. In addition, the nebula can be used to suggest, though not guarantee, that the giant was on the asymptotic giant branch (AGB) rather than on the red giant branch (RGB) at the time of the CE, something that is less certain for other binaries where $M_{\mathrm{c}} \geq 0.47$~\ms \ (see below). All the PN we use are those listed in \citet[][]{DeMarco2011} with the addition of 10 objects: UU~Sge and KV~Vel, V664~Cas and A~65 (as listed by \citealt{Davis2010}), HaTr7 and ESO 330-9 (\citealt{Hillwig2017}), M3-1 \citep{Jones2019}, Hen~2-155 \citep{Jones2015}, Sp~1 \citep{Hillwig2016} and V651~Mon \citep[as recently revised by][note that for this object the companion may not be a main sequence star but instead a subgiant, if so then the secondary would be much closer to filling its Roche radius than we have listed in Table~\ref{tab:obs_demarco}]{Brown2019}.
Finally, we include all the sdO and sdB objects listed by \citet{Davis2010} and most of those listed by \citet{Schreiber2003}, with the exclusion of MT Ser and V477 Lyr. 

The observed quantities in Tables~\ref{tab:obs_zorotovic} and \ref{tab:obs_demarco} derive from observed  lightcurves, sometimes in multiple bands. By modelling these lightcurves one can obtain stellar masses ($M_{\rm c}$ and $M_2$) and orbital separation ($a_{\rm f}$), though more often than not some assumptions and indirect estimates are made. From them, $M_1$ and $R_1$ the mass and radius of the primary giant at the time of CE can be reconstructed using the method of \citet{DeMarco2011}. 

Key to this reconstruction is the determination of whether a given system is post-RGB or post-AGB. If $M_{\rm c} < 0.47$~\ms \ we consider the system to be post-RGB, because masses smaller than this value cannot ignite helium. If the mass is above this value, the star could be a post-AGB or it could be a more massive post-RGB star caught before helium ignition (see discussions in \citealt{Zorotovic2010} and \citealt{DeMarco2011}). However, simply based on the argument that there are a lot more low mass than massive stars, primaries' masses above 0.47~\ms\ are more likely to be those of less massive post-AGB stars rather than those of the rarer, more massive post-RGBs. There are several other sources of uncertainty in the reconstruction method, some quite large, all discussed by \citet{DeMarco2011}, as well as by \citet{Zorotovic2010,Zorotovic2011},  \citet{Davis2010} and \citet{Davis2012}. However, for the current study the evolutionary phase when the CE took place, RGB or AGB, is the most problematic source of uncertainty, so we discuss it further in Section~\ref{ssec:obs_comparison}.

To determine the main sequence mass of the primary we interpolate linearly the initial-to-final mass relation (figure~2 of \citealt{DeMarco2011}), after augmenting the derived core mass, $M_{\rm c}$, by 0.028~\ms, to account for the core growth that would have taken place if the CE had not interrupted the evolution. The primary mass at the time of the CE is then derived from the main sequence value taking into account mass-loss ($M_1/M_{\rm MS}$ = 0.90 for the RGB and $M_1/M_{\rm MS}$ = 0.75 for the AGB). The radius at the time of the CE is then determined using equations 19 and 20 of De Marco et al. (2011).

Some of the observed systems have evolved star masses large enough that it is likely that the main sequence progenitor had a mass in excess of 5~\ms, the upper limit of the initial-to-final mass relation used by \citet{DeMarco2011}. For such objects we determined the mass of the main sequence progenitor by a linear interpolation of the initial-to-final mass relation of \citet{Weidemann2000} (their table~3). The additive term of 0.028~\ms \ to determine the core mass that our stars would have had at the natural end of their giant phases is likely too small for more massive stars, as it was determined using stellar models in the range 0.8-2.5~\ms. As a result, we likely underestimated the main sequence mass and hence also the mass at the time of the CE for these more massive stars. In light of this additional uncertainty, we therefore display those more massive star observations as grey symbols in Figure~\ref{fig:mass_ratio_vs_final_separation}, \ref{fig:binding_energy_vs_final_separation} and \ref{fig:alpha_vs_binding_energy_fit}.

From the values of $M_1$ and $R_1$ we calculate the envelope binding energy at the time of the CE, $E_{\mathrm{bin}}$, as adapted by  \citet{DeMarco2011}:
\begin{equation}
E_{\rm bin} = -\frac{G}{2} \frac{M_{\mathrm{e}} \big( \frac{M_{\mathrm{e}}}{2} + M_{\mathrm{c}} \big)}{\lambda R_1} \ ,
\label{eq:binding_energy}
\end{equation}
where $M_{\mathrm{e}}=M_1 - M_{\rm c}$ is the envelope mass and $\lambda$ is the stellar structure parameter (see, e.g., \citealt{Webbink1984}). The factor of one half is to account for thermal energy via the Virial theorem. This formalism is shown to be accurate by numerical integration of 1D stellar structures. We derive the values of $\lambda$ using the fitting equations of \citet{DeMarco2011}. It should be clarified, however, that these values of $\lambda$ are different from those derived by, e.g., \citet{Loveridge2011} who fitted stellar structure models to a {\it different} analytical representation of the binding energy ($E_{\rm bind} = G M_1 M_{\rm e} / \lambda R_1$, were the variables have the same meaning as in Equation~\ref{eq:binding_energy}). A comparison of their expression with that in Equation~\ref{eq:binding_energy} immediately shows that our values of $\lambda$ are approximately 40\% of those of \citet{Loveridge2011}. This does not affect the values of the binding energies so long as the same analytical approximation as that used to derive the values of $\lambda$ is applied. We list our values in Tables~\ref{tab:obs_zorotovic}, \ref{tab:obs_demarco} and \ref{tab:sim} for clarity and completeness.

The companions' radii were derived using $R_2/R_{\odot} = (M_2/M_{\odot})^{0.8}$ \citep{Torres2010}, which can be shown to be reasonably close to the measured values when such values exist, even accounting for the fact that post-CE companions tend to be larger than their mass would dictate due to irradiation by the primary (\citealt{DeMarco2008}), or by recent accretion during the CE phase \citep{Jones2015}. In at least one case, V651~Mon, \citet{Brown2019} strongly suspects that the secondary is on the subgiant branch and therefore quite a bit larger than our assumption. 
The final separation, $a_{\mathrm{f}}$, is that derived from the observed period, this could be somewhat smaller than the one after the CE event because of magnetic braking (\citealt{Zorotovic2011b}), though not enough to shift the data points in our logarithmic plots.

\subsection{Discussion of the observational sample}
\label{ssec:obs_comparison}  
\begin{figure*}
\includegraphics[scale=0.40, trim=3.5cm 0.0cm 0.0cm 0.0cm]{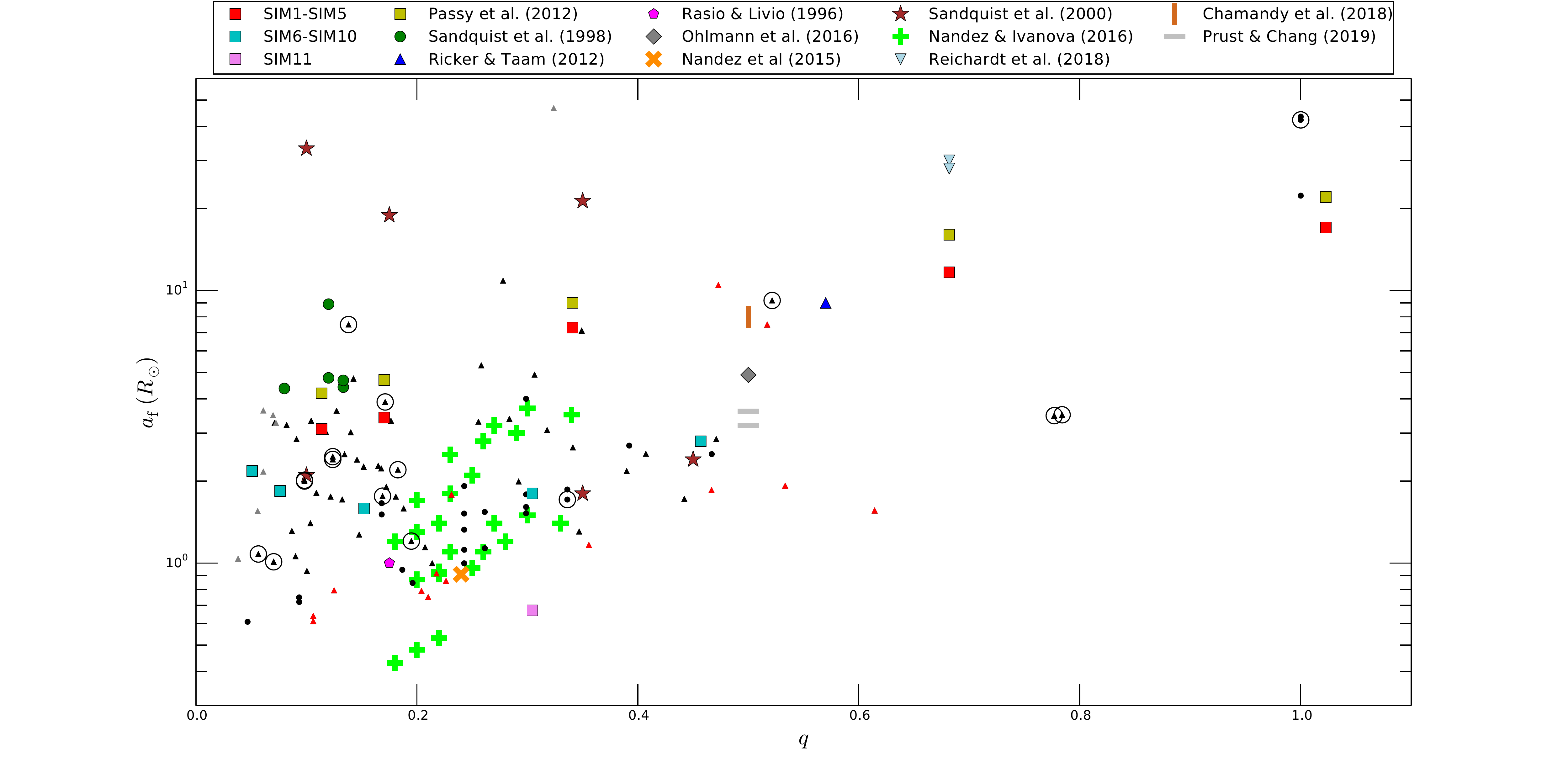}

\caption{\protect\footnotesize{Final orbital separation vs. mass ratio $q=M_2/M_1$. Observed final separations from different authors (Table~\ref{tab:obs_demarco}) and from \citet{Zorotovic2011b} are shown as dots for post-RGB stars and as triangles for post-AGB stars.
In red we mark the observations with $0.47$~\ms~$\leq M_{\mathrm{c}} \leq 0.5$~\ms, in grey those corresponding to primaries with main sequence masses greater than 5~\ms, in black all the remaining ones.
The symbols surrounded by a circle represent systems with PN. 
Simulations from the literature are also shown with coloured symbols, according to the legend. All values of $a_{\rm f}$ should be considered upper limits.}}
\label{fig:mass_ratio_vs_final_separation}
\end{figure*}

In Figure~\ref{fig:mass_ratio_vs_final_separation} we show $a_{\mathrm{f}}$ as a function of $q$, while in Figure~\ref{fig:binding_energy_vs_final_separation} (upper panel) we have plotted $a_{\mathrm{f}}$ as a function of $|E_{\mathrm{bin}}|$ (absolute value of the binding energy). Both these figures are simplistic in that we know the final separation to be dependent on more than one parameter (see Equation~1). However, the structure they reveal is informative. Later (Figure~\ref{fig:alpha_vs_binding_energy_fit}), we will combine quantities in a more complex way. We also include simulation results, which we will discuss in Section~\ref{sec:simulations}.

Observations assumed to derive from RGB primaries are shown as small filled circles and those deriving from AGB stars are shown as small filled triangles. Observations represented by grey symbols denote stars with reconstructed main sequence masses greater than 5~\ms, as explained in Section~\ref{ssec:obs_data}. Circled symbols are post-CE binaries in PN.

We have assumed that all observations with $M_{\mathrm{c}} < 0.47$~\ms \ are post-RGB. In this way we obtain 18 pRGB and 44 pAGB in the \citet{Zorotovic2011b} sample (Table~\ref{tab:obs_zorotovic}) and 6 pRGB and 32 pAGB in the mixed sample (Table~\ref{tab:obs_demarco}). If we assumed a larger limit of  $\leq 0.50$~\ms\ (this value is somewhat arbitrary),  the relative numbers would be 25 pRGB and 37 pAGB in the \citet{Zorotovic2011b} sample and 13 pRGB and 25 pAGB in the mixed sample, therefore  increasing the number of primaries assumed to have undergone a CE interaction on the RGB. 
By increasing the $M_{\mathrm{c}}$ limit all binaries with $0.47$~\ms~$\leq M_{\mathrm{c}} \leq 0.50$~\ms \ move to higher $q$ values in Figure~\ref{fig:mass_ratio_vs_final_separation}, because changing the reconstruction technique from the AGB to RGB methods increases $M_1$ by a factor between 1.5 and 1.8, equally reducing $q$ and increasing their $|E_{\mathrm{bin}}|$ values (Figure~\ref{fig:binding_energy_vs_final_separation}, upper panel). Specifically, the region in Figure~\ref{fig:binding_energy_vs_final_separation} (upper panel) at $|E_{\mathrm{bin}}| < 5 \times 10^{45}$~erg becomes scarcely populated and the affected objects relocate in the zone with $5 \times 10^{45} \leq |E_{\mathrm{bin}}| \leq 3 \times 10^{46}$~erg. While we will adopt the $M_{\mathrm{c}}$ limit proposed by \citet{DeMarco2011}, we have
marked all the points in the range $0.47$~\ms~$\leq M_{\rm c} \leq 0.50$~\ms \ in red as a reminder not to place too much weight on their specific locations in the diagrams. There is no need, however, to track these few data points further, because their being considered post-RGB rather than post-AGB systems does not alter the distribution of post-RGB systems in any systematic way.

The data in Figure~\ref{fig:binding_energy_vs_final_separation} is distributed in three zones: a left zone with $|E_{\mathrm{bin}}| < 3 \times 10^{46}$~erg, that includes only AGB stars; a central one with $3 \times 10^{46} \leq |E_{\mathrm{bin}}| \leq 6 \times 10^{46}$~erg, that includes a mix of RGB and AGB stars and a right zone with $|E_{\mathrm{bin}}| > 6 \times 10^{46}$~erg, that has only RGB stars. 
This segregation is logical, because RGB stars are on average more bound. More bound stars may naturally lead to smaller final separation, something that is only slightly apparent in Figure~\ref{fig:binding_energy_vs_final_separation}, with an absence of systems with $a_{\rm f} \lesssim 10$~\rs, for $|E_{\rm bin}| \gtrsim 5 \times 10^{46}$~erg.

We now consider the objects included in the three zones denoted as ``1"  (low $|E_{\mathrm{bin}}|$, high $a_{\mathrm{f}}$), ``2" (high $|E_{\mathrm{bin}}|$, low $a_{\mathrm{f}}$) and ``3" (low $|E_{\mathrm{bin}}|$, low $a_{\mathrm{f}}$) in Figure~\ref{fig:binding_energy_vs_final_separation} (upper panel; see the data reported in Table~\ref{tab:region_3}). 
Regions 1 and 3 contain eight and seven objects, respectively, each with a similar primary $|E_{\mathrm{bin}}|$ range, but different final separations. Inspecting Table~\ref{tab:region_3} it is clear that the primaries of the two groups are very similar, with $M_1$, $M_{\mathrm{c}}$ and $R_1$ corresponding to average values within the reconstructed sets of observations, although we notice that all the primaries in region 1 and 3 have $M_{\mathrm{c}} > 0.47$~\ms, indicating that the progenitors were more likely on the AGB at the time of CE. The difference between the two zones, is that the binaries in region 1 have more massive companions ($\sim 0.33$~\ms \ on average) than those in region 3 ($\sim 0.11$~\ms \ on average). This suggests that {\it low companion masses lead to small orbital separations}.

Region 2 instead includes three objects that have similar $M_{\mathrm{c}}$ (0.33~\ms \ on average) and the same $M_1$, as they all fall in the RGB reconstruction regime. The reconstruction for these objects yields a low value of $R_1$ (23~\rs \  on average), and hence a high $|E_{\mathrm{bin}}|$, something that may be the cause of such low $a_{\mathrm{f}}$ even if the $M_2$ values of these three systems are relatively high and comparable with those of region 1. 

\begin{table*}
\begin{center}
\begin{adjustbox}{max width=\textwidth}
\begin{tabular}{ccccccccc}
\hline
& Object & $q$ & $M_1$ & $M_{\mathrm{c}}$ & $R_1$ & $M_2$ & $a_{\mathrm{f}}$ & $|E_{\mathrm{bin}}|$ \\
& &  & (\ms) & (\ms) & (\rs) & (\ms) & (\rs) & ($10^{46}$~erg) \\
\hline
\multirow{7}{*}{Region 1}
&     INCMa      & 0.31 & 1.40 & 0.57 & 284 & 0.43 & 4.91  & 1.85 \\
&    Feige24     & 0.28 & 1.40 & 0.57 & 284 & 0.39 & 10.86 & 1.85 \\
& SDSS1519+3536  & 0.14 & 1.40 & 0.57 & 284 & 0.20 & 4.75  & 1.85 \\
& SDSS1646+4223  & 0.26 & 1.01 & 0.53 & 236 & 0.26 & 5.31  & 1.05 \\
& SDSS0924+0024  & 0.35 & 0.92 & 0.52 & 226 & 0.32 & 7.13  & 0.87 \\
& SDSS2318-0935  & 0.52 & 0.73 & 0.50 & 209 & 0.38 & 7.50  & 0.49 \\
& SDSS1434+5335  & 0.47 & 0.68 & 0.49 & 197 & 0.32 & 10.47 & 0.39 \\
& Sp 1 (PN)      & 0.52 & 1.30 & 0.56 & 272 & 0.67 & 9.20  & 1.60 \\
\hline
\multirow{3}{*}{Region 2}
&   CSS080502    & 0.30 & 1.07 & 0.35 & 29 & 0.32 & 1.61 & 14.13 \\
& SDSS1731+6233  & 0.30 & 1.07 & 0.34 & 25 & 0.32 & 1.52 & 16.44 \\
& SDSS2123+0024  & 0.19 & 1.07 & 0.31 & 16 & 0.20 & 0.95 & 26.59 \\
\hline
\multirow{7}{*}{Region 3}
&     NN Ser      & 0.10 & 1.10 & 0.54 & 247 & 0.11 & 0.94 & 1.23 \\
&     XY Sex      & 0.11 & 0.73 & 0.50 & 209 & 0.08 & 0.61 & 0.49 \\
&  BUL-SC 16 335  & 0.22 & 0.73 & 0.50 & 209 & 0.16 & 0.92 & 0.49 \\
&   PG 1017-086   & 0.11 & 0.73 & 0.50 & 198 & 0.08 & 0.64 & 0.52 \\
&     NY Vir      & 0.20 & 0.73 & 0.50 & 198 & 0.15 & 0.79 & 0.52 \\
&  HS 0705+6700   & 0.21 & 0.62 & 0.48 & 181 & 0.13 & 0.75 & 0.30 \\
&    2231+2441    & 0.12 & 0.60 & 0.47 & 171 & 0.07 & 0.79 & 0.29 \\
\hline
\end{tabular}
\end{adjustbox}
\end{center}
\begin{quote}
\caption{\protect\footnotesize{From top to bottom, separated by horizontal lines, main characteristics of Region 1, Region 2, Region 3. As showed in Figure~\ref{fig:binding_energy_vs_final_separation} (upper panel).
Each zone is ordered by decreasing $M_{\mathrm{c}}$.}} \label{tab:region_3}
\end{quote}
\end{table*}

\subsubsection{Post-CE binaries with planetary nebulae}

Most PN (circled points in Figures~\ref{fig:mass_ratio_vs_final_separation} to \ref{fig:alpha_vs_binding_energy_fit}) derive from lower mass AGB primaries. A rare few could derive from RGB progenitors, if their masses are known to be too low to have gone through AGB evolution. In our sample, ESO~330-9 has a very low central stars masses (0.40~\ms). Despite high uncertainties in mass derivations, this PN mass is almost certainly below the helium-burning limit. The mass range derived for this PN central star is 0.38-0.45~\ms\ using all available evidence \citep{Hillwig2017}, and the adopted value of 0.40~\ms\ is arbitrarily chosen to then derive all other other parameters in a consistent way. The PN shape is indistinct as the nebula is large but faint, in line with an RGB origin. V651~Mon (better known by its nebular name, NGC2346) has a mass of 0.46~\ms, right at the limit and with a large uncertainty \citep{Brown2019}. Although we have carried out a post-RGB reconstruction as also advised by \citet{Brown2019}, this PN is distinctly bipolar and looks a lot more like a bona fide PN, throwing some further uncertainty on the quantities reconstructed for this object. Even eliminating these two ``post-RGB" PN, the distribution of the post-AGB PN is not particularly correlated nor concentrated.  

\subsubsection{The widest post-CE binary observations}

Finally, let us consider the four systems that lie at large $a_{\mathrm{f}}$ ($>$20~\rs), namely IK~Peg, V651 Mon, V1379~Aql and FF~Aqr. All four of these objects have main sequence companions more massive than average. The primary masses of V651~Mon, FF~Aqr and V1379~Aql are low enough that we consider them post-RGB. However, applying our post-RGB reconstruction technique results in a primary mass at the time of CE that is smaller than the companion mass, something that cannot be for these systems. For these three systems we therefore assume $M_1 = M_2$, i.e., the smallest primary mass that is consistent with evolutionary theory, which is also the most likely because of the initial mass function (e.g., \citealt[][$M_1$ could in fact be slightly smaller than $M_2$, as some mass loss would have taken place between the main sequence and the CE interaction]{Kroupa2001}). 

\citet{Davis2010} concluded that V651~Mon and FF~Aqr are not the result of a CE interaction during one of the primary's giant phases, because they cannot reconcile their location on the $M_2$ vs. period diagram with the statistical location of systems from their population synthesis models. They favoured instead, based on the same models, a dynamical mass-transfer phase on the stellar thermal time-scales, also known as a thermally-unstable Roche lobe overflow. Even though the V651~Mon system shows the presence of a PN, \citet{Davis2010} and \citet{deKool1993} (as well as \citealt{Brown2019} based on different arguments) argued that it could have formed due to enhanced stellar winds during the thermally unstable Roche lobe overflow. 

The fourth system, IK~Peg, has instead a high $M_{\mathrm{c}} = 1.19$~\ms. Following the AGB reconstruction the AGB progenitor mass is larger than the companion's in accordance with theory. \citet{Landsman1993}, \citet{Smalley1996}, \citet{Davis2010} and \citet{Zorotovic2011b}, agree that the most likely binary interaction for IK~Peg was a CE phase. We also note that IK~Peg is one of 7 systems with a reconstructed main sequence mass larger than 5~\ms, and marked as a grey symbol in Figure~1, 2 and 3, due to the uncertainty on the exact value. The other 6 objects have a similar $|E_{\mathrm{bin}}|$ to IK~Peg, but have much smaller final separations $a_{\rm f}$, likely because of their much lower mass companions. 

FF~Aqr and V1379~Aql listed by \citet{Davis2010} have an additional peculiarity. The small core mass of V1379~Aql, but large implied main sequence primary mass leads to a very small radius at the time of CE and, as a result, a small value of $a_{\rm i}$. This is not consistent with the large final separation of this system, which is larger than the reconstructed initial separation. In the case of FF~Aqr, the delivered orbital energy has a negative value because of relatively little in-spiral ($a_{\rm f} \lesssim a_{\rm i}$), and large $M_1$ ($M_{\rm c} M_2/2a_{\rm f} - M_1 M_2/2a_{\rm i} < 0$). A more complex explanation than the one we use in our reconstruction is likely needed here and as a result we cannot plot these two systems in Figure~\ref{fig:alpha_vs_binding_energy_fit}.

In summary, the four systems with large final separations are likely due to two distinct phenomena. One group suffered a Roche lobe overflow phase that either prevented the CE or reduced its intensity, leaving the binary with a wider separation. The second group (with currently only IK~Peg as member) derives from a regular common envelope, and the large separation has to do with the high mass of the companion. 

\subsubsection{Distinct $\alpha_{\rm CE}$ in post-RGB and post-AGB, post-CE binaries}

In Figure~\ref{fig:alpha_vs_binding_energy_fit} we plot Equation~1, where on the y-axis we plot the binding energy and on the x-axis the difference in orbital energies. In such a plot, the gradient is proportional to $\alpha_{\rm CE}$. Just looking at the observations it is clear that the post-AGB observations, including the PN, stand clear of the rest. Their low binding energy do not translate in a smaller amount of in-spiral (smaller difference between initial and final orbital energies).  The post-RGB observations have larger binding energies, as already discovered, and have a slightly better overlap  with the simulations (which we will discuss in Section~\ref{ssec:sim_comparison}). They too appear uncorrelated. The errors on the location of each observed systems are substantial and thoroughly discussed by \citet{DeMarco2011} and \citet{Zorotovic2010}. This likely explains the scatter. The post-RGB observations overlap the range of $\alpha_{\rm CE}$=0.1-1 with 5 resting at $\alpha_{\rm CE}$ larger than unity, and likely due to the error. The post-AGB observations sit instead at lower values of $\alpha_{\rm CE}$ overall. If those post-AGB objects with masses between 0.47 and 0.50 were instead post-RGB object, there would be fewer objects below the $\alpha_{\rm CE}$=0.1 line. In conclusion, random errors likely explain much of the spread, but we conclude that post-AGB observations, with systematically lower primary binding energies and a similar (albeit wider) range of final separations compared to the RGB sample, point to overall lower values of $\alpha_{\rm CE}$.


\section{Common envelope simulations}
\label{sec:simulations}
In this section we discuss a set of simulations. Most of these simulations have been already collected and compared by \citet{Iaconi2017}. Here we concentrate on their characteristics displayed in the $a_{\mathrm{f}}$ vs. $q$ (Figure~\ref{fig:mass_ratio_vs_final_separation}), $a_{\mathrm{f}}$ vs. $|E_{\rm bin}|$ (Figure~\ref{fig:binding_energy_vs_final_separation}) and $|E_{\rm bin}|$ vs. $\Delta E_{\rm orb}$ (Equation~1; Figure~\ref{fig:alpha_vs_binding_energy_fit}). 

\subsection{The simulation data}
\label{ssec:sim_data}
\citet{Iaconi2017} compared several simulations from the literature and showed that, perhaps surprisingly, different modelling techniques seem to produce reasonably consistent results, at least at the level of precision that we are interested in at this time.
Input and output parameters of the simulations can be found in table~1 of \citet{Iaconi2017}, but we have reproduced it in Table~\ref{tab:sim}, with some updates, for ease of consultation. The values of $R_2$ in Table~\ref{tab:sim} are calculated by assuming that all the companions are main sequence stars. In some of the simulations this cannot be the case as a main sequence companion would overflow its Roche lobe radius. When this is the case we must conclude that the interaction would have resulted in a merger, unless the companion was a much smaller, degenerate object.

\subsubsection{Grid simulations}
\label{sssec:grid}
The grid simulations we consider are those of \citet[][yellow squares in Figure~\ref{fig:mass_ratio_vs_final_separation},  \ref{fig:binding_energy_vs_final_separation}, and Figure~\ref{fig:alpha_vs_binding_energy_fit}, upper panels]{Passy2012} and those carried out with the AMR version of the same code ({\sc Enzo}; \citealt{Bryan2014}) by \citet[][]{Iaconi2018}, for which we will use the original nomenclature SIM1-SIM5 (red squares), SIM6-SIM10 (cyan squares), SIM11 (pink squares) and SIM12 (brown squares).

We also consider the simulations of \citet[][green circles]{Sandquist1998} and \citet[][red stars]{Sandquist2000}, carried out with a static nested grid technique, that of \citet[][blue triangle]{Ricker2012}, carried out with the AMR code {\sc flash} \citep{Fryxell2000} and the recent simulation of \citet[elongated orange rectangle]{Chamandy2018}, carried out with the code {\sc AstroBear} \citep{Carroll-Nellenback2013}. This simulation is not listed by Iaconi et al. (2017). It models a 1.9~\ms\ RGB star with a radius $R_1 = 48$~\rs, and a companion with a 0.98~\ms\ mass placed at an initial separation of $a=49$~\rs\ and achieving a final separations of $a_{\rm f} = 8$~\rs. Their inner resolution is in the range 0.07-0.14~\rs.

We do not consider the high initial eccentricity simulations of \citet{Staff2016a} which are the only ones including high mass (3~\ms) AGB stars with relatively massive companions (up to $q=1$). The final separation is by far the largest seen in simulation, $\sim 200$~\rs, but this may be due to a very coarse resolution. 

\subsubsection{Smooth particle hydrodynamics simulations}
\label{sssec:sph}
\begin{figure*}
\centering     
\includegraphics[scale=0.40, trim=6.3cm 0.0cm 6.3cm 3.0cm]{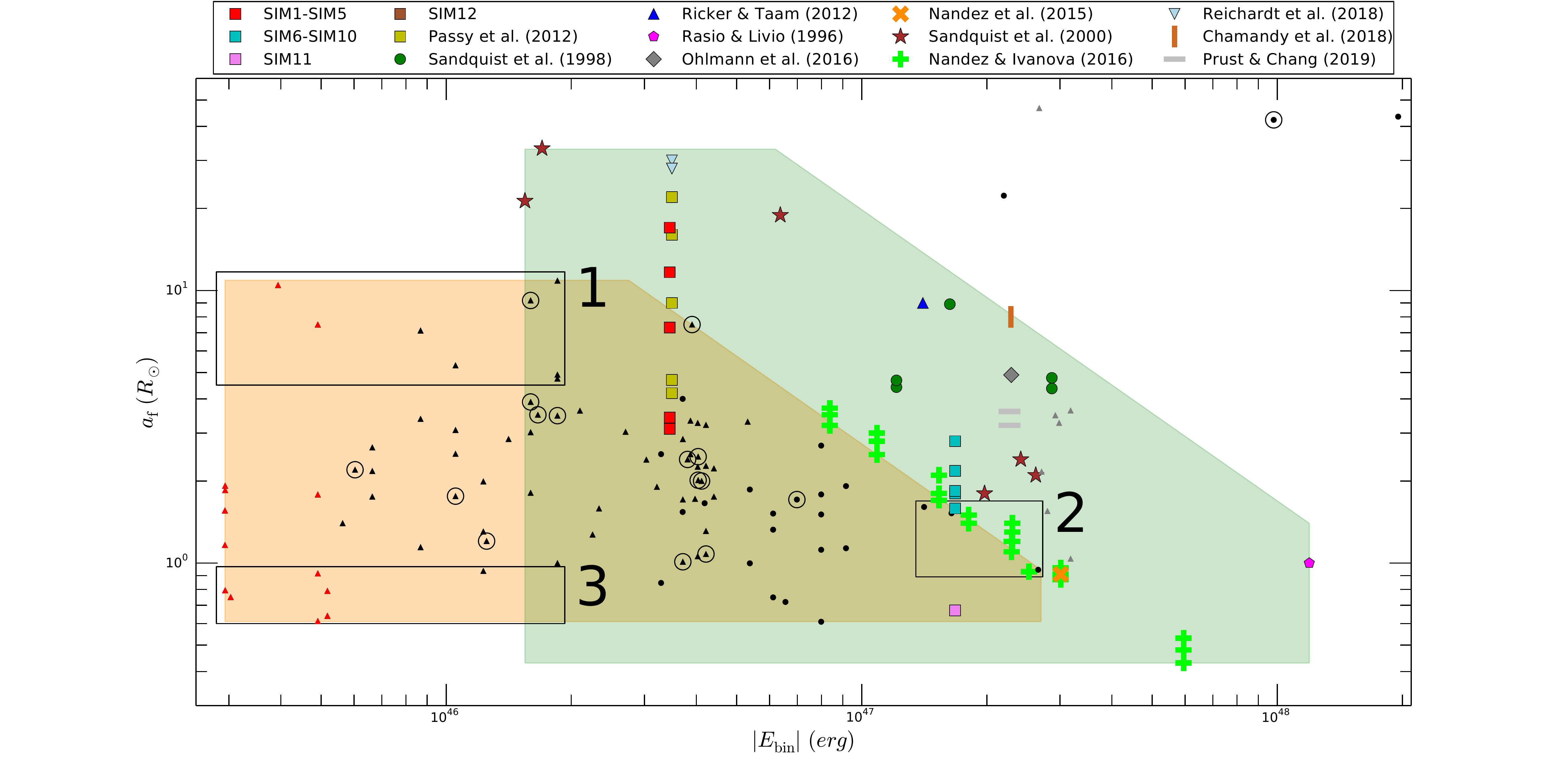}
\includegraphics[scale=0.40, trim=6.3cm 0.0cm 6.3cm 0.0cm]{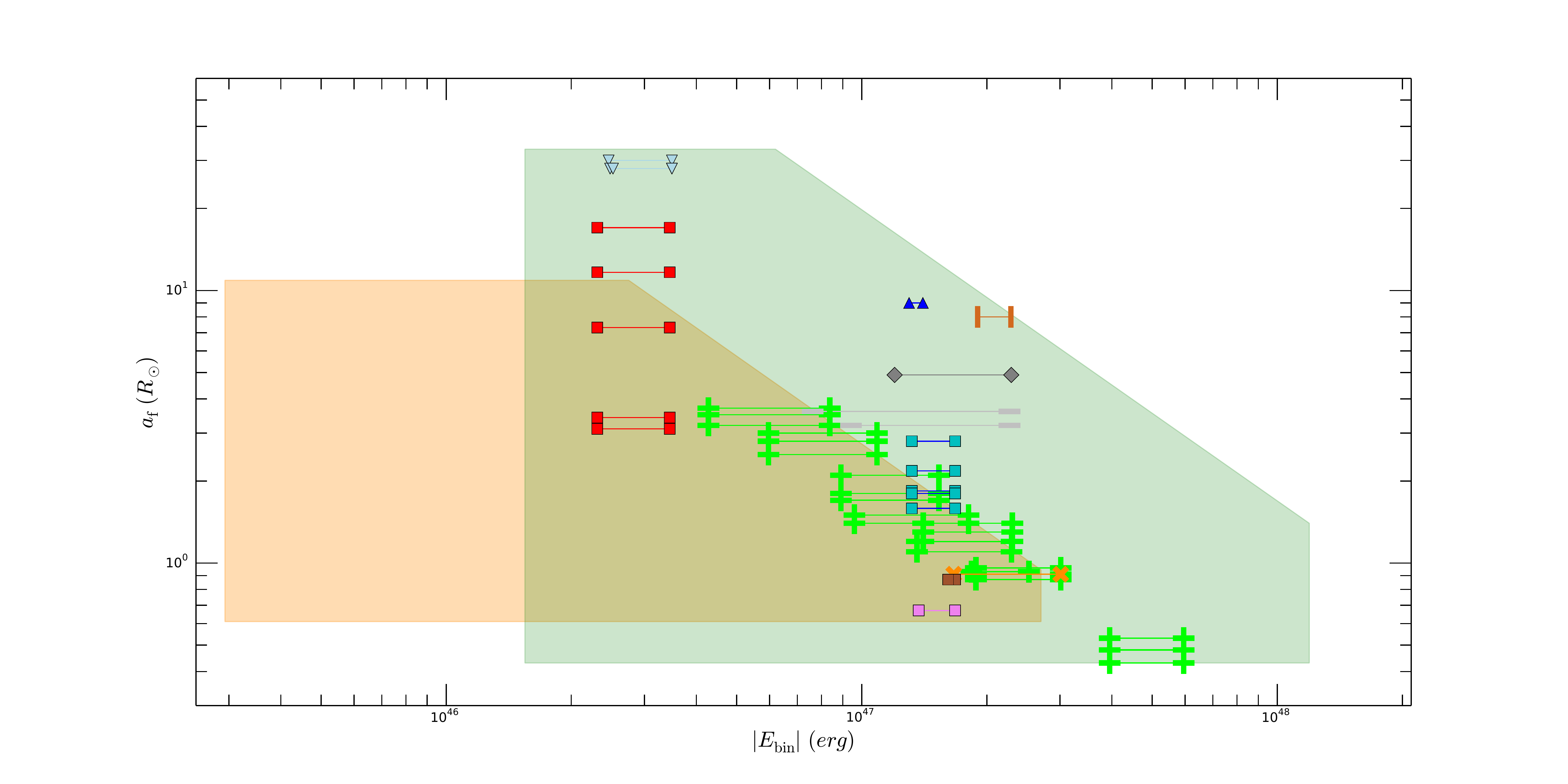}

\caption{\protect\footnotesize{Upper panel: $a_{\rm f}$ vs. $|E_{\mathrm{bin}}|$ of the primary at the time of the CE. Observed systems are represented as done in Figure~\ref{fig:mass_ratio_vs_final_separation}.
Simulations from the literature are shown with coloured symbols, according to the legend.
All the binding energies are computed by using equation~13 of \citet{DeMarco2011}. The orange-shaded region represents the area occupied by observations. The green-shaded region encloses all the simulations and its sides are parallel to those of the orange-shaded. The numbered rectangles represent the samples discussed in the text.
Lower panel: initial binding energy of simulated primary stars. For each simulation the left point is obtained by using the numerical binding energies, while the right one is calculated with 
Equation~13 of \citet{DeMarco2011}. If the same simulation is carried out at different resolutions we mark the corresponding points with different colours. The shaded areas, for comparison, are the same as for the upper panel.}}
\label{fig:binding_energy_vs_final_separation}
\end{figure*}

We also list simulations carried out with the SPH code {\sc phantom} \citep{Price2018} by \citet[][with increasing resolutions 80\,000, 230\,000 and 1.1 million SPH particles; light blue triangles in Figures~\ref{fig:mass_ratio_vs_final_separation}, \ref{fig:binding_energy_vs_final_separation} and \ref{fig:alpha_vs_binding_energy_fit}]{Reichardt2019}, where a simulation with the same parameters of the grid simulation in \citet{Iaconi2017} is started at a wider initial separation so that the entire unstable Roche lobe overflow is simulated. 

The SPH simulations from the literature are those of \citet[][pink pentagon]{Rasio1996}, \citet[][orange cross]{Nandez2015},  and \citet[][green plusses, see table~1 of \citet{Iaconi2017} and Table~\ref{tab:sim} for the simulations' parameters]{Nandez2016}. The simulations of \citet{Nandez2015} and \citet{Nandez2016} are carried out with a tabulated equation of state.

\subsubsection{Moving mesh simulations}
\label{sssec:moving-mesh}
The {\sc arepo} moving mesh simulations of \citet[][grey diamond; see also \citealt{Ohlmann2016b,Ohlmann2017}]{Ohlmann2016}, brought into the field a new method, a hybrid between grid and Lagrangian method. 

While this paper was in review, a new moving mesh simulations of the common envelope interaction appeared in the literature. \citet{Prust2019} used a new moving mesh code called {\sc manga} \citep{Chang2017}, to simulate a 2~\ms, 52~\rs, RGB star with a core of 0.38~\ms. The initial separation of the simulation was such that the 1\ms\ companion was placed at the surface of the giant. This simulations have the same parameters as those of \citet{Ohlmann2016}. Two simulations were carried out: a non-rotating one and one that was placed in 95\% corotation. For each of these two simulations, an ideal gas equation of star and a tabulated equation of state were used. The fraction of unbound envelope was $\sim$10\% for the ideal gas equation of state simulation (Prust, private communication), irrespective of the envelope rotation, while most $\sim$65\% of the envelope was unbound for the tabulated equation of state simulations (though more unbinding would presumably take place if the simulation had been allowed to run for more than 240~days). The final separations of these 4 simulations ranged between 3.2 and 3.6~\rs\ (Table~\ref{tab:sim}).

\subsubsection{Determination of the final separation and of the giant star's binding energy in simulations}
\label{sssec:af-ebin}
Values of $a_{\mathrm{f}}$ plotted in  Figure~\ref{fig:mass_ratio_vs_final_separation} and \ref{fig:binding_energy_vs_final_separation} and used in Figure~\ref{fig:alpha_vs_binding_energy_fit} are obtained using the criterion of \citet{Sandquist1998}, namely, the separation at the time when the orbital shrinking timescale becomes one tenth of its maximum value (see \citealt{Iaconi2018} for a more accurate description), for all simulations where this was available (see table~1 of \citealt{Iaconi2017}), while for the rest of the simulations we utilised the values listed by the respective authors. These values would likely decrease if the simulations were run for longer, with some values decreasing more than others. The timescales for the further reduction of these values would be longer than those sampled so far, and likely allow some dynamical readjustment of the stellar structure. We here point out that irrespective of the definition of $a_{\rm f}$, these values are likely upper limits to the actual final separation, the one that would be achieved in Nature.

In Figure~\ref{fig:binding_energy_vs_final_separation} (upper panel), the values of $|E_{\mathrm{bin}}|$ are obtained analytically applying Equation~\ref{eq:binding_energy} to the simulation data from the respective papers, similarly to what was done with observations. The stellar structure parameter $\lambda$ is calculated using the formulation of \citet{DeMarco2011} for RGB stars, their equation~17. In Figure~\ref{fig:binding_energy_vs_final_separation}, 
lower panel, we show the same area as in upper panel, but display also the values of $|E_{\mathrm{bin}}|$ calculated  numerically, by integrating the simulation data when available. We comment on the comparison between these two methods of determination of the binding energy in Section~\ref{sssec:Ebin}. In Figure~\ref{fig:alpha_vs_binding_energy_fit} we use the numerical $|E_{\mathrm{bin}}|$ where available and the analytical where not (we place a circle around those simulations for which the analytical expression for the binding energy had to be used).

\subsection{Discussion of the simulated sample}
\label{ssec:sim_comparison}
Here we discuss the simulations as a group using Figures~\ref{fig:mass_ratio_vs_final_separation} to \ref{fig:alpha_vs_binding_energy_fit}. Simulations are tabulated in Table~\ref{tab:sim}.

\subsubsection{The binding energy of simulated giant stars}
\label{sssec:Ebin}
In Figure~\ref{fig:binding_energy_vs_final_separation} (lower panel) we observe a shift between the value of $|E_{\mathrm{bin}}|$ calculated analytically (larger values) and the numerically integrated values for those simulations where these values are reported (smaller values; values are connected by a bar forming a dumbbell). The numerical values are always obtained adding the potential and thermal energies of the envelopes. We believe the shift to be mainly an effect of resolution: for SIM9, SIM11 and SIM12, which are a resolution sequence, SIM12, with a highest innermost resolution of 0.05~\rs, has a numerical $|E_{\mathrm{bin}}|$ which is almost exactly the same as the one derived analytically, while SIM9, with the lowest resolution (0.84~\rs) has the lowest value of the numerical $|E_{\mathrm{bin}}|$ and most discrepant with the analytical value (see also the discussion in appendix~A2 of \citealt{Iaconi2018}). At higher grid resolution stars are mapped with more accuracy near their central region and the resulting value of the binding energy is higher and  more similar to the analytical value.   

The simulation of \citet{Ricker2012} and that of \citet{Chamandy2018} have $|E_{\mathrm{bin}}|$ values that are quite close to the analytical values. This is likely due to the high resolution of those simulations: 0.29~\rs\ and $\sim$0.15~\rs. This is in line with the behaviour we observe for SIM11 and SIM12 of \citet{Iaconi2018} that have resolutions of 0.21 and 0.05~\rs, respectively.

The three {\sc phantom} simulations by \citet{Reichardt2019}, with three different resolutions (80\,000, 230\,000 and 1.1 million SPH particles, respectively) do not have very different binding energy values. This is likely due to the lower factor of 1.2 and 1.8 improvement in linear resolution between simulation pairs (compared with a factor of 4 improvement in linear resolution for pairs SIM9/SIM11 and SIM11/SIM12)

Perhaps unexpectedly, the moving mesh simulations by \citet{Ohlmann2016} and \citet{Prust2019} show a large gap between the numerical and analytical values of $|E_{\mathrm{bin}}|$, even though they have high resolutions (0.01~\rs \ and 0.1~\rs, respectively). We do not understand why such high resolution simulations would have such discrepancy. 

In summary, application of the analytic formula results in a larger value of  $|E_{\mathrm{bin}}|$. Simulations with a higher resolution, seem to have numerical values of $|E_{\mathrm{bin}}|$ that are larger and closer to the analytic value, but this may not be universally true and may depend on initial parameters and on the specifics of the code used. Since the behaviour of the simulation is a function of the binding energy of the envelope {\it in the simulation}, in the following discussion we use the numerically-derived value of the binding energy, where available, and we indicate every time when this is not the case.

\subsubsection{A comment on the dependence of the final separation on simulation resolution and on softening length}
\label{sssec:resolution}
As we have discussed in Section~\ref{sssec:af-ebin} one can consider the simulated values of $a_{\rm f}$ to be upper limits in all cases because of a range of reasons. Here we discuss specifically two factors that can artificially limit the in-spiral: the spatial resolution and the size of the softening length applied to the gravitational potential of the particles.

Spatial resolution tests have been performed, e.g., by \citet{Sandquist1998}, \citet{Passy2012}, \citet{Ohlmann2016PhD}, \citet{Staff2016b}, \citet{Iaconi2018} and \citet{Reichardt2019}. They are unanimous in concluding that higher resolutions result in smaller $a_{\rm f}$, although the extent of the reduction of $a_{\rm f}$ is variable. Among the main reasons for this behaviour is that more resolved stars have larger $|E_{\mathrm{bin}}|$, causing more in-spiral (Section~\ref{sssec:Ebin}) and that higher spatial resolution usually also means a smaller  softening length that increases the strength of the interaction between particles and gas and leads to more in-spiral.

The softening length size can affect the strength of the interaction under certain circumstances \citep{Sandquist1998}. It can also halt the in-spiral once the orbital separation is similar to the softening length. This means that when the published $a_{\rm f}$ is of the order of the softening length we must consider it an upper limit. This is the case in SIM1, SIM2 and SIM6-SIM11 of \citet{Iaconi2018} and possibly in some of the simulations of \citet{Nandez2015} and \citet{Nandez2016}.

Combined resolution and softening length tests have been carried out by some authors. Notable examples are those by \citet{Sandquist1998} and \citet{Chamandy2019}. In the first case reducing the softening length resulted in a smaller  $a_{\rm f}$, while in the second case in its increase. However, these two results are hardly comparable. The former study compared a constant softening length simulation to one where they reduced the softening length gradually. Both final separations were much larger than the softening length, but the simulation with a decreasing softening length resulted in a smaller separation, indicating that they were probing the effects of softening length size on the interaction {\it strength}. In the latter study, softening lengths and resolutions were changed by a set amount late in the in-spiral allowing them to probe differences over a small part of the interaction, and the difference in final separations and other quantities were very small indeed. 

{\it The cascade effects of the interplay of resolution and softening length are extremely difficult to quantify and will require more exhaustive tests in the future.}


\subsubsection{Considerations on the energy formalism from the standpoint of simulations}
\label{ssec:alpha-sims}
In Figure~\ref{fig:alpha_vs_binding_energy_fit}, we plot the left and right-hand sides of Equation~1 for all the simulations, in such a way that any gradient represents $\alpha_{\rm CE}$. We utilise the numerical values of the binding energy, where available, and analytical ones where not (the symbols have in those cases been circled). Over-plotted are lines of constant $\alpha_{\rm CE}$ from 0.1 to 1.0.

As described in Section~\ref{sssec:af-ebin} it is likely that {\it all} simulations' final separations are upper limits. Some are definitely upper limits, those where $a_{\rm f}$ is similar to the softening length, those where it is shown that a higher resolution decreases $a_{\rm f}$, or those that only simulate a handful of orbits. For the others the value may be very close to what it should be. It is not useful, however, to argue  the exact extent to which each simulated value of $a_{\rm f}$ is an upper limit. Therefore we assume all simulated value of $a_{\rm f}$ to be upper limits in Figure~\ref{fig:alpha_vs_binding_energy_fit}.

Most simulations do not unbind the envelope, leaving open the question of what will the future evolution of the system and of the orbit look like. However, we note that the simulations of \citet{Nandez2016} indicate that the final separation is decided before the envelope is fully unbound by the action of recombination energy. Further, Reichardt et al. (in preparation) showed that both the simulations of \citet{Passy2012} and of \citet{Nandez2016} have very similar final separations when performed with an ideal gas or a tabulated equation of state, meaning that the envelope unbinding at the hand of recombination energy does not impact the final separation substantially.

In Figure~\ref{fig:alpha_vs_binding_energy_fit} the simulations of \citet{Nandez2016} are well correlated. The $\alpha_{\rm CE}$ values predicted in this way ($\alpha_{\rm CE} \lesssim 0.6$) are similar, though slightly smaller to those listed by \citet[][$0.87-1.0$]{Nandez2016}. The neat distribution of the \citet{Nandez2016} simulations on our plot does point to the fact that the lifting of the gas out of the orbital volume (even before it is unbound) is closely related to its total binding energy, $E_{\rm bin}$, even if that binding energy is not reduced to zero by the in-spiral itself. There could therefore be a physical connection between envelope weight and its impact on the early lifting of that weight, which determines the final separation of the binary. This relation could also hold for those simulations that do not unbind the envelope, but that may, like the simulations by \citet{Nandez2016}, unbind it later on, by using a source of energy that does not impact the separation.

If this relationship could be expressed as the $\alpha$-formalism of Equation~\ref{eq:alpha_formalism}, that would be helpful, even if it would not have the same physical meaning envisaged originally, but rather $\alpha_{\rm CE}$ would give the proportionality constant between the total envelope binding energy and the orbital energy needed for the separation to stabilise\footnote{Analytically, we know that the separation stabilises when the gravitational friction becomes zero, which happens either because of the envelope gas co-rotating with the orbital motion, or because of the evacuation of envelope gas from the orbit, or a combination of both. In simulations, however, there is more variability as gas flows are complex and 3-dimensional, and strong density gradients exist (analytical theory does not account for those) and the entire simulated interaction is mediated by the softening length, whose effects are ill-quantified - see discussion in \citet{Reichardt2019}.}.

In Figure~\ref{fig:alpha_vs_binding_energy_fit} all simulations that reside on the right of the $\alpha_{\rm CE} =1$ line, can provide some constraints. These are the simulations of \citet[][$\alpha_{\rm CE} <1$ to $\alpha_{\rm CE} <0.6$]{Nandez2016} and \citet[][$\alpha_{\rm CE} <0.8$]{Nandez2015}, and those of \citet[][$\alpha_{\rm CE} <1$]{Iaconi2017}, \citet[][SIM11; $\alpha_{\rm CE} <0.2$]{Iaconi2018} and \citet[][$\alpha_{\rm CE} <0.6$]{Prust2019}. {\it These simulations tell us that $\alpha_{\rm CE}$ is unlikely to be a constant factor.} We notice in passing that for some of these simulations the final separation is so small that if the companion is a main sequence star it would overflow its Roche lobe (Table~\ref{tab:sim}). This means that for these simulations a main sequence companion results in a merger, while an evolved companion would survive.

Moreover comparing the location of the simulations of \citet{Iaconi2017} and \citet{Reichardt2019} in Figure~\ref{fig:alpha_vs_binding_energy_fit}, which have all been carried out with the same primary star, tell us that $\alpha_{\rm CE}$ is likely dependent also on what happened before the fast in-spiral. This conclusion is supported by the in-depth analysis carried out by \citet[][see their section 4.4]{Iaconi2017} who showed that side by side comparisons in both grid and SPH simulations, where the only difference was the initial separation, result in wider final separations for wider initial separations. They also concluded that this is due to the primary being allowed to expand before the in-spiral starts, rather than due to the extra angular momentum imparted to the primary.

For completeness we point out that the values of $\alpha_{\rm CE}$ reported by Sandquist et al. (1998) were calculated in a different way: they determined what part of the envelope was unbound in their simulations and what the binding energy of that part of the envelope was, by knowing where that gas resided before the interaction. This gave them the ejection efficiency for the gas that was truly unbound. They found that approximately half of the orbital energy had unbound the envelope while half has propelled that part of the envelope to a certain kinetic energy above what was needed to unbind it. It is not clear whether they did account for the amount of orbital energy that lifted but did not unbind the remainder of the envelope gas. In any case this is a very different way to calculate $\alpha_{\rm CE}$ and the values they obtain should not be compared with those discussed here.


\section{Comparison between observations and simulations}
\label{sec:obs_sim_comparison}

\begin{landscape}
\begin{figure}
\includegraphics[scale=0.6, trim=5.0cm 0.0cm 0.0cm 0.0cm]{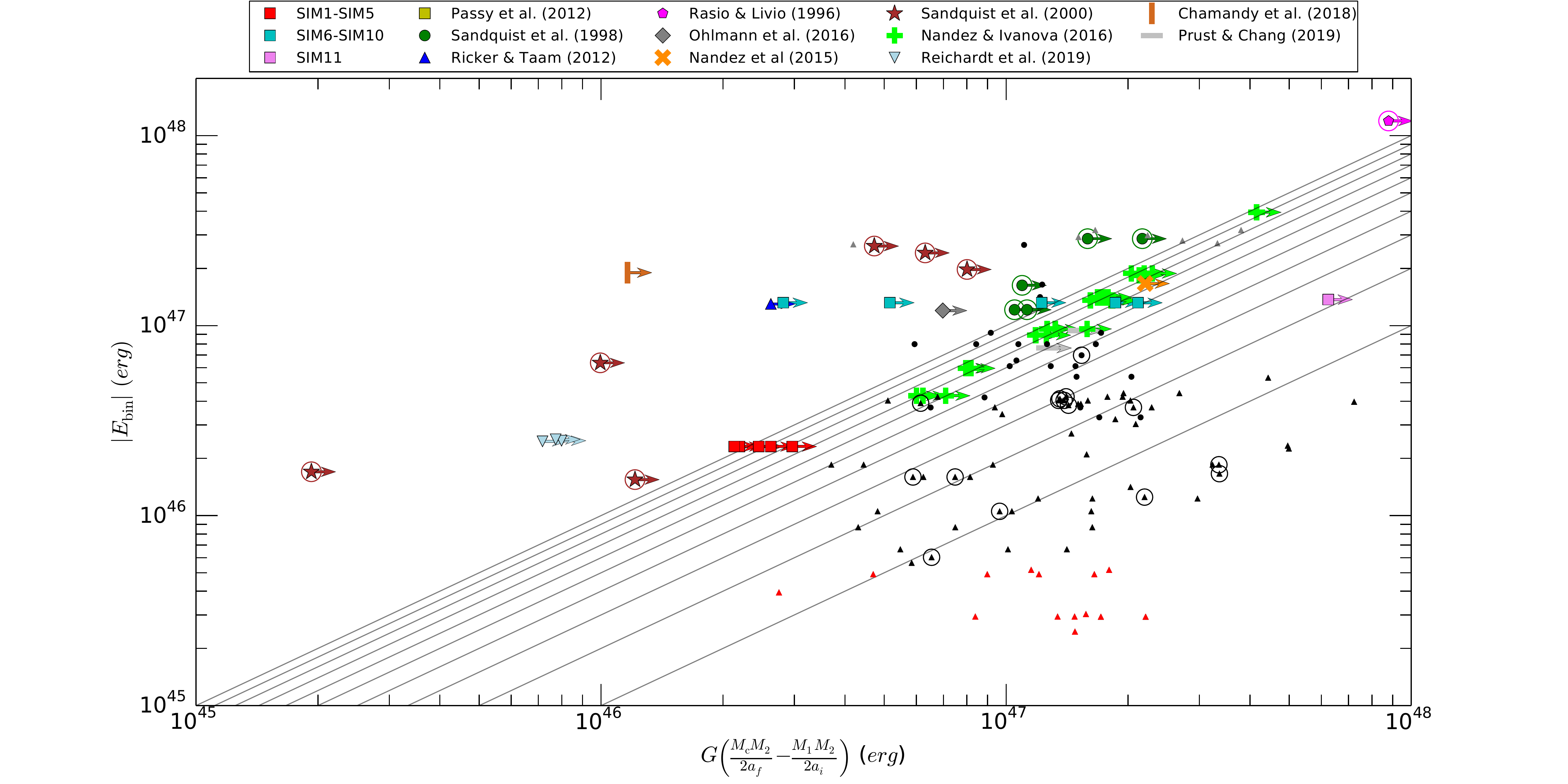}
\caption{\protect\footnotesize{$|E_{\mathrm{bin}}|$ vs. $\Delta E_{\rm orb}$ for the same systems shown in Figure~\ref{fig:mass_ratio_vs_final_separation} and \ref{fig:binding_energy_vs_final_separation}. Here, for simulations where it was available, we used the numerical value of $|E_{\mathrm{bin}}|$, the simulations for which we were not able to retrieve it are shown with a circle around their respective mark. The diagonal lines represent, from top to bottom, lines corresponding to $\alpha = $~1.0, 0.9, 0.8, 0.7, 0.6, 0.5, 0.4, 0.3, 0.2 and 0.1. The arrows on the simulated points are meant to remind the reader that all the simulations' values for $a_{\rm f}$ are upper limits.}}
\label{fig:alpha_vs_binding_energy_fit}
\end{figure}
\end{landscape}

Finally, in this section we compare the distributions of observations and simulations on all planes (Figure~\ref{fig:mass_ratio_vs_final_separation}, \ref{fig:binding_energy_vs_final_separation} and \ref{fig:alpha_vs_binding_energy_fit}. All observational data are in Tables~\ref{tab:obs_zorotovic} and \ref{tab:obs_demarco}, while simulation data are in Table~\ref{tab:sim}).

Figure~\ref{fig:mass_ratio_vs_final_separation} shows a reasonable overlap of the locus of simulations and observations. Common to both distributions is a dearth of systems with low separation and high mass ratio. These systems would be easy to find so the paucity of observations in that parameter space is not an observational bias. Though only few simulations exist in this space, it does appear the case that the higher the mass ratio the larger the separation.

There is an offset between the locus of observations and simulations, highlighted by the orange and green areas in Figure~\ref{fig:binding_energy_vs_final_separation}. 
We do not have reason to believe that the observations have systematically too low a value of $|E_{\mathrm{bin}}|$ or too low a value of the orbital separation, except that some of the red triangles could shift to the right if we increased the helium-burning mass limit (Section~\ref{ssec:obs_comparison}). 

The lack of simulations at low $|E_{\mathrm{bin}}|$ is simply explained by the fact that simulations tend to model compact RGB stars, which naturally have relatively larger values of $|E_{\mathrm{bin}}|$. It is more difficult to stabilise the naturally unstable AGB stars in the 3D computational domain; their larger sizes and low density envelopes also make for longer computational times. As a result research has naturally progressed more rapidly in the RGB domain.

Yet, even accounting for this, we still see that the simulations have larger $|E_{\mathrm{bin}}|$ than the high $|E_{\mathrm{bin}}|$, post-RGB observations. The shift between analytical and numerical values of $|E_{\mathrm{bin}}|$ discussed in Section~\ref{sssec:resolution} (Figure~\ref{fig:binding_energy_vs_final_separation} - lower panel) is partly to blame. The numerical (smaller) values of $|E_{\rm bin}|$ are what is actually ``felt'' by the gas in the simulation, and it is therefore what actually determines the outcome of the simulated in-spiral and the final separation. For a proper comparison we ought to use the numerical, not the analytical $|E_{\mathrm{bin}}|$, which would shift the simulations to the left, resulting in a better overlap. 

As we have described in Section~\ref{sssec:af-ebin} we need to recognise that for a range of different reasons all simulated values of $a_{\rm f}$ are upper limits. Shifting the simulations down in Figure~\ref{fig:binding_energy_vs_final_separation} would also result in closing the gap between simulations and observations.

In Figure~\ref{fig:alpha_vs_binding_energy_fit} we see that the post-RGB observations have larger values of $|E_{\mathrm{bin}}|$ and approximately overlap with the simulations. The spread of the observations is almost certainly mostly due to the large errors so one may not expect any trends to be visible. At best we can only conclude that the average location of the post-RGB observations is similar to those simulations that have a similar $\Delta E_{\rm orb}$ range.  

This is not so for the post-AGB observations. They too can be expected to have large error bars, but they are located on average at positions in Figure~\ref{fig:alpha_vs_binding_energy_fit} that have lower binding energies than the post-RGB observations and the simulations for similar $\Delta E_{\rm orb}$. Even excluding completely the data with primary core masses near the limit of 0.47~\ms\ (red triangles), which have the most extreme values, does not substantially alter this conclusion.

\subsection{Wide post-CE binaries in observations and simulations}
\label{ssec:the_post-CE_binaries_with_the_largest_separations}

What leads to a common envelope interaction ending with a relatively wide separation, wider than typically observed in post-CE objects? A light envelope, such as that of an upper AGB star, leading to a swift unbinding and large final separation is not the answer, because many post-AGB observations, including those inside PN, are detected at small final separations. 
What does the comparison of observations and simulations suggest?

There are 4 {\it observations} (IK Peg, V1379~Aql, FF~Aqr and V651~Mon) with final separations in excess of 20~\rs. Three are post-RGB observations and have high values of $q \sim 1$. The fourth object with $a_{\rm f} > 20$~\rs\ is a massive post-AGB (IK~Peg; Figure~\ref{fig:mass_ratio_vs_final_separation}) with a much lower value of $q$. In Figure~\ref{fig:binding_energy_vs_final_separation} we see that these observations have relatively large $|E_{\mathrm{bin}}|$.
All three post-RGB observations with high $a_{\rm f}$ cannot be placed in Figure~\ref{fig:alpha_vs_binding_energy_fit}, because the x-axis value is negative (owing to very little in-spiral from a presumed starting separation as well as a relatively massive $M_1$). The post-AGB, massive, IK~Peg is above all trend lines, but this could be due to a large error bar. As discussed in Section~\ref{ssec:obs_comparison} three of the observations with $a_{\rm f} > 20$~\rs\ may derive from a prolonged phase of Roche lobe overflow, possibly followed by an anomalously weak CE or no CE at all.

There are 4 distinct {\it simulations} with final separations in excess of 10~\rs: the simulation of \citet{Reichardt2019}, two simulations of \citet{Iaconi2018} and one of \citet[][we are not counting two of the three high $a_{\rm f}$ simulations of \citealt{Sandquist2000}, that are certain upper limits; Section~\ref{sssec:resolution}]{Sandquist2000}. 
We note that on Figures~\ref{fig:mass_ratio_vs_final_separation} and \ref{fig:binding_energy_vs_final_separation} these simulations have the full range of $q$ values and intermediate values of $|E_{\mathrm{bin}}|$. In this regard, these four simulations do not shed light on the behaviour of the observations described above.

This said, the simulation of \citet{Reichardt2019} may suggest an explanation for  the behaviour of the observations: with the exact same primary and companion stars as SIM4 of \citet[]{Iaconi2018}, but starting with a wider separation, it did not achieve as much in-spiral and stands out from the group in the region of Figure~\ref{fig:alpha_vs_binding_energy_fit} where $\alpha_{\rm CE} > 1$. \citet{Reichardt2019} suggested that simulating the Roche lobe overflow phase leads to a wider separation. They also remarked that analytical theory \citep[e.g.,][]{Tout1991} predicts that a larger $q$  promotes a longer and more stable Roche lobe mass transfer phase, so they suggested that simulating larger $q$ values than they had may lead to even larger separations than they obtained, and maybe this could explain the observations of post-RGB systems with high $q$ and large $a_{\rm f}$. To test these suggestions, initial simulations with increasing values of $q$ and that also simulate the Roche lobe overflow phase, were carried out. Early results (De Marco et al. in preparation) show a trend of larger $a_{\rm f}$ with larger $q$ values, all other parameters remaining the same. Resolution, always a problem in these simulations, can be shown not to affect the final separation, but the effect of resolution on the length and stability of the Roche lobe overflow phase remains ill quantified.

Understanding the impact of the Roche lobe overflow phase on the outcome of CE simulations is important for many reasons. If it were needed it would seriously impact simulation times. In the case of the simulations of \citet{Reichardt2019} including the pre-CE, Roche lobe overflow phase shifts the value of $\Delta E_{\rm orb}$ away from the observations in Figure~\ref{fig:alpha_vs_binding_energy_fit}, something that is not easy to interpret. 

\subsection{Close post-CE binaries in observations and simulations}
\label{ssec:the_post-CE_binaries_with_the_smallest_separations}
Finally, we take a look at those simulations and observations with very small values of the final separation. The frequency and type of such systems may reveal what causes CE interactions to result in a merger.

The thirteen smallest $a_{\rm f}$ values ($a_{\rm f } < 1$~\rs) in the {\it observational sample} belong to eight post-AGB and five post-RGB binaries. They all cluster at low $q$ values (Figure~\ref{fig:mass_ratio_vs_final_separation}) and have $M_2<0.22$~\ms. All of these systems are close to overflowing their Roche lobe with some clearly achieving Roche lobe overflow. Since there is no evidence in the literature that any of these post-CE binaries are in contact, we must conclude that values of $R_2/R_{\rm RL,2} > 1$ in Table~\ref{tab:obs_zorotovic} and \ref{tab:obs_demarco} are due to measurement error. Values of $|E_{\mathrm{bin}}|$ are spread over almost the entire range (Figure~\ref{fig:binding_energy_vs_final_separation}) with the post-AGB systems naturally having smaller values. We note however, that all the post-AGB systems except for one are red triangles, indicating that they are close to the boundary that divides the post-RGB from the post-AGB systems, so they could be post-RGB, in which case their $|E_{\mathrm{bin}}|$ would be larger, and there would be a better correlation between very bound envelopes and very small separations. Even eliminating the red triangles, we see that in Figure~\ref{fig:alpha_vs_binding_energy_fit} the post-AGB systems, with small values of  $|E_{\mathrm{bin}}|$, have very small separations implying very low $\alpha_{\rm CE}$ values. In fact the 4 systems with $\Delta E_{\rm orb} > 3.5 \times 10^{47}$~erg are all post-AGB systems, have low values of $q < 0.26$ and relatively low values of $\alpha_{\rm CE}$. The question is therefore whether AGB systems have  a systematically different common envelope behaviour, represented by a lower value of $\alpha_{\rm CE}$, or whether the scatter is simply due to measurement error. 

Nine simulations have similarly low separations and even lower if they are upper limits. Eight of them are those of \citet{Ivanova2016} and \citet{Nandez2015}. If the companions in these simulations were assumed to be main sequence stars (instead of white dwarfs, as assumed in their paper), most of these simulations would be overflowing their Roche lobes, likely implying that a binary like those simulated would result in a merger. The ninth simulation is the high resolution SIM11 of \citet[][pink square in Figures~\ref{fig:mass_ratio_vs_final_separation}-\ref{fig:alpha_vs_binding_energy_fit}]{Iaconi2018}. Its very low final separation implies $\alpha_{\rm CE} < 0.2$. We should note, however, that in this simulation the secondary would be grossly overflowing its Roche lobe, indicating that such a system would only survive if the secondary were a white dwarf. Such low value of $\alpha_{\rm CE}$ may only apply to such double degenerate systems\footnote{The next smallest final separation, at 1.0~\rs, is that of \citet{Rasio1996} and is explained by also having the highest value of $|E_{\mathrm{bin}}|$ simulation of all (Figure~\ref{fig:alpha_vs_binding_energy_fit}).}.

The post-RGB star distribution overlaps the bulk of the \citet{Nandez2016} simulations (with the scatter of the distribution likely dictated by uncertainty), though there are no post-RGB observations on the rightmost side of Figure~\ref{fig:alpha_vs_binding_energy_fit}, indicating that when the companions are main sequence stars (as is the case for our selected observations) such small separations lead to a merger. 

All of these systems have such small orbits that the companions are close to filling their Roche lobe (see Table~\ref{tab:obs_zorotovic} and \ref{tab:obs_demarco}, where the values of $R_2/R_{\rm RL,2}$ are all around unity). {\it This would indicate that there are many similar binaries that have merged in the common envelope}. We wonder  where the disrupted companion gas would settle (in the core or in the envelope) and what behaviour this type of merger would trigger \citep{Nordhaus2006}. The existence of such low $a_{\rm f}$, low $M_2$ systems may also have implications for the formation of CVs: many of these systems would form CVs with very low mass companions immediately after the CE. These systems would bypass the period gap altogether \citep{Politano2004}.


\section{Summary and conclusions}
\label{sec:conclusions}
We have carried out a comparative analysis to understand how observations of post-CE systems and the population of systems resulting from 3D hydrodynamic simulations relate to each other. The analysis is carried out by using the final separation, $a_{\mathrm{f}}$, vs. mass ratio ($q = M_2 / M_1$, where $M_1$ is the primary mass and $M_2$ is the companion mass) plane, the $a_{\mathrm{f}}$ vs. $|E_{\mathrm{bin}}|$ plane, where $|E_{\mathrm{bin}}|$ is the absolute value of the binding energy of the primary at the moment of the common envelope interaction and the $|E_{\mathrm{bin}}|$ vs. $ \Delta E_{\rm orb}$ plane, for which the gradient is $\alpha_{\rm CE}$.

To compare observations and simulations we have reconstructed the properties of the primary stars at the moment of CE using parameters determined from observations of the post-CE systems along with the method of \citet{DeMarco2011}. The uncertainties in this reconstruction method are large and introduce a significant source of noise in our discussion. Despite these uncertainties, the main conclusions of this comparison have been:

\begin{enumerate}

\item We consider all {\it simulated} final separations  upper limits even though some are likely much closer to the actual value than others. This means that the implied values of $\alpha_{\rm CE}$ are upper limits too. We could expect that the simulations of \citet{Nandez2015} and \citet{Nandez2016}, that unbind the entire envelope, provide us with values of $a_{\rm f}$ that are closer to Nature. However, our analysis gives us no reason to believe that the inclusion of recombination energy leading to full envelope removal leads to systematically different values of the final separation, at least for the modelled parameter space. 
The simulations of \citet{Nandez2015} and \citet{Nandez2016} and a few others that do not unbind the envelope, agree on $\alpha_{\rm CE} \lesssim 0.6-1.0$, depending on the simulation, with some of that variability possibly real. The \citet{Nandez2015} and \citet{Nandez2016} simulations alone may indicate lower values of $\alpha_{\rm CE}$ for less bound stars, in agreement with point (ii) below. The existence of a simulation (Iaconi et al. 2018) indicating $\alpha_{\rm CE} < 0.2$ also points to a range of $\alpha_{\rm CE}$ values, though the very small $a_{\rm f}$ value of this simulation makes it plausible only if the companion is a white dwarf.

\item The distribution of post-RGB {\it observations} approximately overlaps the trend of the \citet{Nandez2016} simulations, with the scatter easily explained by the large error bars. The observed post-AGB systems (including the PN) systematically  reside at lower values of $\alpha_{\rm CE}$.  There are not enough simulations of AGB CEs, and in particular of low mass AGB progenitors, to help with an explanation. {\it AGB common envelopes appear to have a very different behaviour from RGB interactions. We predict that simulations of AGB common envelopes will reveal lower values of $\alpha_{\rm CE}$.}

\item {\it Observations and simulations} point to the fact that low mass companions (and to an extent low $q$ values) result in small final separations, but more so for the post-RGB sample. The tightest post-CE binaries in our observed sample have companion masses as low as $\sim$0.05~\ms, and are all very close to filling their Roche lobe. This  indicates that $0.05-0.1$~\ms\ is the the actual smallest companion mass that survives CE rather than being due to observational bias. The post-AGB observed sample (considering in particular those post-AGB binaries with $M_{\rm c} > 0.50$~\ms) have larger separations for lower mass companions, compared to the post-RGB sample and the separation distribution is very broad. Very few simulations have been carried out with low mass companions ($M_2 < 0.2$) and some have relatively large final separations. However, the upper limit nature of these separations does not allow us to draw conclusions in this regard. 

\item High $q$ {\it observations} with large separations can be presumed to have gone through a prolonged phase of Roche lobe overflow. {\it Simulations} cannot yet demonstrate the same behaviour but they suggest that simulating the Roche lobe overflow phase, may lead to wider separations (\citealt{Reichardt2019}). If this were the case,
it is possible that interactions where the pre-CE phase is relatively important may not be described by an $\alpha$-type formalism. Similar simulations with larger values of $q$ should be attempted to determine the relationship between the intensity of the Roche lobe overflow phase and the outcome of the subsequent CE. 

\item {\it The issue of the dependence of simulations' final separation with resolution is pressing.} A low resolution results in a lower value of  $|E_{\mathrm{bin}}|$ and a larger value of the final separation. It is also clear that the value of the smoothing (softening) length contributes to the uncertainty: it physically limits the in-spiral to the size of the smoothing length. A smaller softening length, however, may also decrease the value of the $a_{\rm f}$ even if the core and companion never approach each other, as thoroughly discussed by \citet{Sandquist2000}. The extent of these effects is not yet well quantified. 
\end{enumerate}

In conclusion, this is a pretty complex landscape. The body of simulations is becoming sizeable, but comparability remains an issue with many subtle definition differences even in simple quantities like the binding energy. We advocate a far greater degree of coordination  and, ultimately, that the simulation data should be publicly available. As for the observations, we wish for the establishment of a single database of observationally-derived parameters of post-CE binaries, alongside the best estimate of their uncertainty as well as homogeneously reconstructed parameters for the systems at the time of the CE interaction.


\section*{Acknowledgments}
\label{sec:acknowledgments}
RI is grateful for the financial support provided by the Postodoctoral Research Fellowship of the Japan Society for the Promotion of Science (JSPS P18753) and by the International Macquarie University Research Excellence Scholarship. OD acknowledges financial support through the Australian Research Council Future Fellowship scheme (FT130100034). We are grateful to David Jones for spotting some omissions, and providing some updates for some of the data presented in Table~\ref{tab:obs_demarco}. We thank Logan Prust, Luke Chamandy and Thomas Reichardt for help with interpreting their simulations and an anonymous referee for a stern review.


\bibliographystyle{aa}
\bibliography{bibliography}{}
\bsp


\appendix
\section{Data derived from observations and simulations}
\label{sec:observations_data}
In this appendix we list data derived from observations and simulations. Table~\ref{tab:obs_zorotovic} is almost identical to the list presented by \citet{Zorotovic2011b}, while in Table~\ref{tab:obs_demarco} we present a series of mixed observations, whose sources are reported in the last column of the table itself. In Table~\ref{tab:sim} we report simulation quantities.

\begin{table*}
\begin{center}
\begin{adjustbox}{max width=0.89 \textwidth}
\begin{tabular}{+c^c^c^c^c^c^c^c^c^c^c^c}
\hline
Object          & $q$  & $M_1$ & $M_{\mathrm{c}}$ & $R_1$ & $\lambda$ & $M_2$ & $R_2$ & $R_2/R_{\mathrm{RL,2}}$ & $a_{\mathrm{f}}$ & $|E_{\mathrm{bin}}|$ &  $G \left( \frac{M_{\rm c}M_2}{2 a_{\rm f}} - \frac{M_1 M_2}{2 a_{\rm i}} \right)$ \\
                &      & (\ms) & (\ms)            & (\rs) &           & (\ms) & (\rs) &                         & (\rs)            & ($10^{46}$~erg)      & ($10^{46}$~erg)                                                                    \\
\hline                                                      
 SDSS0052-0053  & 0.06 & 5.2   & 1.2              & 198   & 0.46      & 0.32  & 0.40  & 1.0                     & 2.2              & 27                   & 33                                                                                 \\
     IKPeg      & 0.32 & 5.2   & 1.2              & 201   & 0.46      & 1.7   & 1.4   & 0.10                    & 47               & 27                   & 4.2                                                                                \\
 SDSS1429+5759  & 0.07 & 5.2   & 1.1              & 182   & 0.46      & 0.38  & 0.46  & 0.75                    & 3.3              & 30                   & 22                                                                                 \\
 SDSS1558+2642  & 0.06 & 5.2   & 1.1              & 171   & 0.46      & 0.32  & 0.40  & 0.62                    & 3.6              & 32                   & 17                                                                                 \\
 SDSS2112+1014  & 0.04 & 5.2   & 1.1              & 171   & 0.46      & 0.20  & 0.28  & 1.7                     & 1.0              & 32                   & 38                                                                                 \\
 SDSS2339-0020  & 0.07 & 4.6   & 0.93             & 151   & 0.43      & 0.32  & 0.40  & 0.62                    & 3.5              & 29                   & 15                                                                                 \\
 SDSS0303+0054  & 0.06 & 4.5   & 0.92             & 152   & 0.43      & 0.25  & 0.33  & 1.2                     & 1.6              & 28                   & 27                                                                                 \\
 SDSS0246+0041  & 0.13 & 3.0   & 0.80             & 1027  & 0.36      & 0.38  & 0.46  & 0.57                    & 3.6              & 2.1                  & 16                                                                                 \\
 SDSS1524+5040  & 0.12 & 2.7   & 0.73             & 683   & 0.35      & 0.32  & 0.40  & 0.61                    & 3.0              & 2.7                  & 14                                                                                 \\
 SDSS2114-0103  & 0.15 & 2.6   & 0.70             & 567   & 0.35      & 0.38  & 0.46  & 0.84                    & 2.4              & 3.0                  & 21                                                                                 \\
 SDSS1414-0132  & 0.10 & 2.5   & 0.67             & 466   & 0.34      & 0.26  & 0.34  & 0.49                    & 3.3              & 3.4                  & 9.8                                                                                \\
     UZSex      & 0.09 & 2.4   & 0.65             & 408   & 0.34      & 0.22  & 0.30  & 0.52                    & 2.9              & 3.7                  & 9.4                                                                                \\
     EGUMa      & 0.18 & 2.4   & 0.64             & 380   & 0.34      & 0.42  & 0.50  & 0.62                    & 3.3              & 3.9                  & 15                                                                                 \\
 SDSS2120-0058  & 0.13 & 2.4   & 0.64             & 380   & 0.34      & 0.32  & 0.40  & 0.71                    & 2.5              & 3.9                  & 15                                                                                 \\
  EC14329-1625  & 0.16 & 2.3   & 0.62             & 330   & 0.34      & 0.38  & 0.46  & 0.85                    & 2.3              & 4.2                  & 19                                                                                 \\
 SDSS1548+4057  & 0.09 & 2.3   & 0.62             & 330   & 0.34      & 0.20  & 0.28  & 1.1                     & 1.3              & 4.2                  & 18                                                                                 \\
  EC12477-1738  & 0.17 & 2.3   & 0.61             & 307   & 0.33      & 0.38  & 0.46  & 0.87                    & 2.2              & 4.4                  & 19                                                                                 \\
  REJ1016-0520  & 0.07 & 2.1   & 0.60             & 294   & 0.33      & 0.15  & 0.22  & 0.36                    & 3.3              & 4.0                  & 5.1                                                                                \\
 SDSS1143+0009  & 0.15 & 2.1   & 0.60             & 294   & 0.33      & 0.32  & 0.40  & 0.76                    & 2.3              & 4.0                  & 16                                                                                 \\
 SDSS1348+1834  & 0.17 & 1.9   & 0.59             & 290   & 0.32      & 0.32  & 0.40  & 0.87                    & 1.9              & 3.2                  & 19                                                                                 \\
     INCMa      & 0.31 & 1.4   & 0.57             & 284   & 0.30      & 0.43  & 0.51  & 0.37                    & 4.9              & 1.9                  & 9.2                                                                                \\
    Feige24     & 0.28 & 1.4   & 0.57             & 284   & 0.30      & 0.39  & 0.47  & 0.16                    & 10               & 1.9                  & 3.7                                                                                \\
 SDSS1519+3536  & 0.14 & 1.4   & 0.57             & 284   & 0.30      & 0.20  & 0.28  & 0.25                    & 4.8              & 1.9                  & 4.5                                                                                \\
  REJ2013+4002  & 0.14 & 1.3   & 0.56             & 272   & 0.29      & 0.18  & 0.25  & 0.37                    & 3.0              & 1.6                  & 6.2                                                                                \\
 RXJ2130.6+4710 & 0.47 & 1.2   & 0.55             & 259   & 0.29      & 0.56  & 0.63  & 0.70                    & 2.9              & 1.4                  & 20                                                                                 \\
     NN Ser     & 0.10 & 1.1   & 0.54             & 247   & 0.28      & 0.11  & 0.17  & 0.88                    & 0.9              & 1.2                  & 12                                                                                 \\
 SDSS0833+0702  & 0.29 & 1.1   & 0.54             & 247   & 0.28      & 0.32  & 0.40  & 0.72                    & 2.0              & 1.2                  & 16                                                                                 \\
 SDSS1411+1028  & 0.35 & 1.1   & 0.54$\pm$0.08    & 247   & 0.28      & 0.38  & 0.46  & 1.2                    & 1.3              & 1.2                  & 30                                                                                 \\
     DE Cvn     & 0.41 & 1.0   & 0.53             & 236   & 0.28      & 0.41  & 0.49  & 0.64                    & 2.5              & 1.1                  & 16                                                                                 \\
 SDSS1646+4223  & 0.26 & 1.0   & 0.53             & 236   & 0.28      & 0.26  & 0.34  & 0.24                    & 5.3              & 1.1                  & 4.8                                                                                \\
 SDSS1718+6101  & 0.32 & 1.0   & 0.53             & 236   & 0.28      & 0.32  & 0.40  & 0.46                    & 3.1              & 1.1                  & 10                                                                                 \\
     LTT560     & 0.21 & 0.92  & 0.52$\pm$0.12    & 226   & 0.28      & 0.19  & 0.26  & 0.91                    & 1.1              & 0.87                 & 16                                                                                 \\
 SDSS0924+0024  & 0.35 & 0.92  & 0.52             & 226   & 0.28      & 0.32  & 0.40  & 0.19                    & 7.1              & 0.87                 & 4.3                                                                                \\
 SDSS1705+2109  & 0.28 & 0.92  & 0.52             & 226   & 0.28      & 0.26  & 0.34  & 0.36                    & 3.4              & 0.87                 & 7.5                                                                                \\
     GKVir      & 0.12 & 0.82  & 0.51             & 217   & 0.27      & 0.10  & 0.16  & 0.41                    & 1.8              & 0.66                 & 5.5                                                                                \\
      HZ9       & 0.34 & 0.82  & 0.51$\pm$0.10    & 217   & 0.27      & 0.28  & 0.36  & 0.47                    & 2.7              & 0.66                 & 10.1                                                                               \\
 SDSS0949+0322  & 0.39 & 0.82  & 0.51$\pm$0.08    & 217   & 0.27      & 0.32  & 0.40  & 0.61                    & 2.2              & 0.66                 & 14.1                                                                               \\
    BPM6502     & 0.23 & 0.73  & 0.50$\pm$0.05    & 209   & 0.27      & 0.17  & 0.24  & 0.52                    & 1.8              & 0.49                 & 9.0                                                                                \\
 SDSS2318-0935  & 0.52 & 0.73  & 0.50$\pm$0.05    & 209   & 0.27      & 0.38  & 0.46  & 0.19                    & 7.5              & 0.49                 & 4.7                                                                                \\
 SDSS1434+5335  & 0.47 & 0.68  & 0.49             & 197   & 0.27      & 0.32  & 0.40  & 0.12                    & 10               & 0.39                 & 2.8                                                                                \\
     MS Peg     & 0.36 & 0.62  & 0.48$\pm$0.02    & 187   & 0.26      & 0.22  & 0.30  & 0.87                    & 1.2              & 0.29                 & 17                                                                                 \\
 SDSS0238-0005  & 0.61 & 0.62  & 0.48$\pm$0.15    & 187   & 0.26      & 0.38  & 0.46  & 0.88                    & 1.6              & 0.29                 & 22                                                                                 \\
 SDSS0110+1326  & 0.53 & 0.60  & 0.47$\pm$0.02    & 171   & 0.26      & 0.32  & 0.40  & 0.64                    & 1.9              & 0.29                 & 15                                                                                 \\
 SDSS1212-0123  & 0.47 & 0.60  & 0.47$\pm$0.01    & 171   & 0.26      & 0.28  & 0.36  & 0.62                    & 1.9              & 0.29                 & 13                                                                                 \\
  EC13349-3237  & 0.47 & 1.1   & 0.46             & 112   & 0.24      & 0.50  & 0.57  & 0.73                    & 2.5              & 3.3                  & 17                                                                                 \\
     LMCom      & 0.26 & 1.1   & 0.45             & 101   & 0.24      & 0.28  & 0.36  & 0.87                    & 1.5              & 3.7                  & 15                                                                                 \\
 SDSS1506-0120  & 0.30 & 1.1   & 0.45             & 101   & 0.24      & 0.32  & 0.40  & 0.36                    & 4.0              & 3.7                  & 6.5                                                                                \\
     RRCae      & 0.17 & 1.1   & 0.44             &  90   & 0.24      & 0.18  & 0.25  & 0.64                    & 1.7              & 4.2                  & 8.8                                                                                \\
 SDSS1435+3733  & 0.24 & 1.1   & 0.42             &  71   & 0.24      & 0.26  & 0.34  & 1.3                     & 1.0              & 5.4                  & 20                                                                                 \\
 SDSS1724+5620  & 0.34 & 1.1   & 0.42             &  71   & 0.24      & 0.36  & 0.44  & 0.82                    & 1.9              & 5.4                  & 15                                                                                 \\
     GD448      & 0.09 & 1.1   & 0.41             &  63   & 0.24      & 0.10  & 0.16  & 1.0                     & 0.8              & 6.1                  & 10                                                                                 \\
 SDSS2216+0102  & 0.24 & 1.1   & 0.41             &  63   & 0.24      & 0.26  & 0.34  & 0.97                    & 1.3              & 6.1                  & 15                                                                                 \\
 SDSS2240-0935  & 0.24 & 1.1   & 0.41             &  63   & 0.24      & 0.26  & 0.34  & 0.84                    & 1.5              & 6.1                  & 13                                                                                 \\
  WD0137-3457   & 0.05 & 1.1   & 0.39             &  49   & 0.24      & 0.05  & 0.09  & 0.91                    & 0.6              & 8.0                  & 5.9                                                                                \\
     CCCet      & 0.17 & 1.1   & 0.39             &  49   & 0.24      & 0.18  & 0.25  & 0.70                    & 1.5              & 8.0                  & 8.4                                                                                \\
     UXCVn      & 0.39 & 1.1   & 0.39             &  49   & 0.24      & 0.42  & 0.50  & 0.61                    & 2.7              & 8.0                  & 11                                                                                 \\
 SDSS1529+0020  & 0.24 & 1.1   & 0.39             &  49   & 0.24      & 0.26  & 0.34  & 1.2                     & 1.1              & 8.0                  & 17                                                                                 \\
 SDSS2132+0031  & 0.30 & 1.1   & 0.39             &  49   & 0.24      & 0.32  & 0.40  & 0.80                    & 1.8              & 8.0                  & 13                                                                                 \\
 SDSS1047+0523  & 0.24 & 1.1   & 0.38             &  43   & 0.24      & 0.26  & 0.34  & 0.67                    & 1.9              & 9.2                  & 9.2                                                                                \\
   CSS080502    & 0.30 & 1.1   & 0.35             &  29   & 0.24      & 0.32  & 0.40  & 0.89                    & 1.6              & 14                   & 12                                                                                 \\
 SDSS1731+6233  & 0.30 & 1.1   & 0.34             &  25   & 0.24      & 0.32  & 0.40  & 0.94                    & 1.5              & 16                   & 12                                                                                 \\
 SDSS2123+0024  & 0.19 & 1.1   & 0.31             &  16   & 0.24      & 0.20  & 0.28  & 1.2                     & 0.95             & 27                   & 11                                                                                 \\
\hline
\end{tabular}
\end{adjustbox}
\end{center}
 \begin{quote}
  \caption{\protect\footnotesize{Observed and derived quantities from \citet{Zorotovic2011b} ordered by decreasing $M_{\mathrm{c}}$. We only list error bars for those values of $M_{\mathrm{c}}$ where the error would shift the value across the limit $M_{\mathrm{c}} = 0.47$~\ms\ separating post-RGB ($M_{\mathrm{c}}<0.47$~\ms) from post-AGB ones ($M_{\mathrm{c}} \geq 0.47$~\ms).}} \label{tab:obs_zorotovic}
 \end{quote}
\end{table*}

\begin{table*}
\begin{center}
\begin{adjustbox}{max width=\textwidth}
\begin{tabular}{+c^c^c^c^c^c^c^c^c^c^c^c^c}
\hline
Object           & $q$  & $M_1$ & $M_{\mathrm{c}}$ & $R_1$ & $\lambda$ & $M_2$ & $R_2$ & $R_2/R_{\mathrm{RL,2}}$ & $a_{\mathrm{f}}$  & $|E_{\mathrm{bin}}| $ & $G \left( \frac{M_{\rm c}M_2}{2 a_{\rm f}} - \frac{M_1 M_2}{2 a_{\rm i}} \right)$ & Reference               \\
                 &      & (\ms) & (\ms)            & (\rs) &           & (\ms) & (\rs) &                         & (\rs)             & ($10^{46}$~erg)       & ($10^{46}$~erg)                                                                   &                         \\
\hline                                                        
    V471 Tau     & 0.26 & 3.6   & 0.84             & 570   & 0.39      & 0.93  & 0.94  & 1.06                    & 3.3               & 5.3                   & 44                                                                                & \citealt{Schreiber2003} \\
     QS Vir      & 0.15 & 2.9   & 0.78             & 918   & 0.36      & 0.43  & 0.51  & 1.73                    & 1.3               & 2.3                   & 50                                                                                &   \citealt{Davis2012}   \\
    BPM 71214    & 0.19 & 2.9   & 0.77             & 866   & 0.36      & 0.54  & 0.61  & 1.56                    & 1.6               & 2.3                   & 49                                                                                &   \citealt{Davis2012}   \\
   BE Uma (PN)   & 0.14 & 2.6   & 0.70             & 440   & 0.35      & 0.36  & 0.44  & 0.26                    & 7.5               & 3.9                   & 6.1                                                                               &  \citealt{DeMarco2008}  \\
 SDSS J0314-0111 & 0.13 & 2.4   & 0.65             & 408   & 0.34      & 0.32  & 0.40  & 1.05                    & 1.7               & 3.7                   & 23                                                                                &   \citealt{Davis2010}   \\
    M3-1 (PN)    & 0.07 & 2.4   & 0.65             & 408   & 0.34      & 0.17  & 0.24  & 1.29                    & 1.0               & 3.7                   & 22                                                                                &   \citealt{Jones2019}   \\
    KOI-3278     & 0.44 & 2.4   & 0.63             & 365   & 0.34      & 1.0   & 1.0   & 1.93                    & 1.7               & 4.0                   & 72                                                                                & \citealt{Zorotovic2014} \\
    A 63 (PN)    & 0.12 & 2.3   & 0.63             & 376   & 0.34      & 0.29  & 0.37  & 0.70                    & 2.4               & 3.8                   & 14                                                                                &  \citealt{DeMarco2008}  \\
    DS 1 (PN)    & 0.10 & 2.3   & 0.63             & 348   & 0.34      & 0.23  & 0.31  & 0.75                    & 2.0               & 4.1                   & 13                                                                                & \citealt{Hilditch1996}  \\
   UU Sge (PN)   & 0.12 & 2.3   & 0.63             & 355   & 0.34      & 0.29  & 0.37  & 0.69                    & 2.5               & 4.0                   & 13                                                                                &   \citealt{Davis2010}   \\
   KV Vel (PN)   & 0.10 & 2.3   & 0.63             & 355   & 0.34      & 0.23  & 0.31  & 0.74                    & 2.0               & 4.0                   & 13                                                                                &   \citealt{Davis2010}   \\
    2009+6216    & 0.08 & 2.3   & 0.62             & 330   & 0.33      & 0.19  & 0.26  & 0.42                    & 3.2               & 4.2                   & 6.8                                                                               &   \citealt{Davis2012}   \\
 Hen 2-155 (PN)  & 0.06 & 2.3   & 0.62             & 330   & 0.33      & 0.13  & 0.20  & 1.04                    & 1.1               & 4.2                   & 14                                                                                &   \citealt{Jones2015}   \\
    1857+5144    & 0.18 & 2.3   & 0.61             & 307   & 0.33      & 0.41  & 0.49  & 1.14                    & 1.8               & 4.4                   & 27                                                                                &   \citealt{Davis2012}   \\
 SDSS J1151-0007 & 0.09 & 2.1   & 0.60             & 294   & 0.33      & 0.19  & 0.26  & 1.25                    & 1.1               & 4.0                   & 20                                                                                &   \citealt{Davis2010}   \\
   HFG 1 (PN)    & 0.78 & 1.4   & 0.57             & 316   & 0.30      & 1.1   & 1.1   & 0.86                    & 3.5               & 1.7                   & 34                                                                                &  \citealt{DeMarco2008}  \\
    2237+8154    & 0.21 & 1.4   & 0.57             & 284   & 0.30      & 0.30  & 0.38  & 1.49                    & 1.0               & 1.9                   & 32                                                                                &   \citealt{Davis2010}   \\
  V664 Cas (PN)  & 0.78 & 1.4   & 0.57             & 284   & 0.30      & 1.1   & 1.1   & 0.86                    & 3.5               & 1.8                   & 34                                                                                &   \citealt{Davis2010}   \\
  NGC 6337 (PN)  & 0.19 & 1.3   & 0.56             & 346   & 0.29      & 0.25  & 0.33  & 1.10                    & 1.2               & 1.2                   & 22                                                                                & \citealt{Hillwig2016b}  \\
    A 65 (PN)    & 0.17 & 1.3   & 0.56             & 272   & 0.29      & 0.22  & 0.30  & 0.32                    & 3.9               & 1.6                   & 5.9                                                                               &  \citealt{Hillwig2015}  \\
    1042-6902    & 0.11 & 1.3   & 0.56             & 272   & 0.29      & 0.14  & 0.21  & 0.54                    & 1.8               & 1.6                   & 8.1                                                                               &   \citealt{Davis2012}   \\
    Sp 1 (PN)    & 0.52 & 1.3   & 0.56             & 272   & 0.29      & 0.67  & 0.73  & 0.24                    & 9.2               & 1.6                   & 7.5                                                                               &  \citealt{Hillwig2016}  \\
   HaTr 7 (PN)   & 0.17 & 1.0   & 0.53             & 236   & 0.28      & 0.17  & 0.24  & 0.57                    & 1.8               & 1.1                   & 9.6                                                                               &  \citealt{Hillwig2017}  \\
    A 46 (PN)    & 0.18 & 0.8   & 0.51$\pm$0.07    & 239   & 0.27      & 0.15  & 0.22  & 0.41                    & 2.2               & 0.60                  & 6.5                                                                               &  \citealt{DeMarco2008}  \\
     AA Dor      & 0.10 & 0.8   & 0.51$\pm$0.12    & 256   & 0.27      & 0.09  & 0.14  & 0.48                    & 1.4               & 0.56                  & 5.8                                                                               &  \citealt{Mueller2010}  \\
   PG 1017-086   & 0.11 & 0.7   & 0.50             & 198   & 0.27      & 0.08  & 0.13  & 0.96                    & 0.64              & 0.52                  & 12                                                                                & \citealt{Schreiber2003} \\
     NY Vir      & 0.20 & 0.7   & 0.50             & 198   & 0.27      & 0.15  & 0.22  & 1.10                    & 0.79              & 0.52                  & 18                                                                                & \citealt{Schreiber2003} \\
  BUL-SC 16 335  & 0.22 & 0.7   & 0.50             & 209   & 0.27      & 0.16  & 0.23  & 0.98                    & 0.92              & 0.49                  & 17                                                                                &   \citealt{Davis2010}   \\
     XY Sex      & 0.11 & 0.7   & 0.50             & 209   & 0.27      & 0.08  & 0.13  & 1.01                    & 0.61              & 0.49                  & 12                                                                                &   \citealt{Davis2010}   \\
     HW Vir      & 0.23 & 0.6   & 0.48$\pm$0.09    & 224   & 0.26      & 0.14  & 0.21  & 0.93                    & 0.86              & 0.24                  & 15                                                                                & \citealt{Schreiber2003} \\
  HS 0705+6700   & 0.21 & 0.6   & 0.48             & 181   & 0.26      & 0.13  & 0.20  & 1.02                    & 0.75              & 0.30                  & 16                                                                                & \citealt{Schreiber2003} \\
    2231+2441    & 0.12 & 0.6   & 0.47             & 171   & 0.26      & 0.07  & 0.13  & 0.72                    & 0.79              & 0.29                  & 8.4                                                                               &   \citealt{Davis2010}   \\
   J2020+0437    & 0.20 & 1.1   & 0.46             & 112   & 0.24      & 0.21  & 0.29  & 1.36                    & 0.85              & 3.3                   & 21                                                                                &   \citealt{Davis2010}   \\
  V651 Mon (PN)  & 1.00 & 3.5   & 0.46             & 59    & 0.20      & 3.5   & 2.7   & 0.17                    & 42                & 98                    & --                                                                                &   \citealt{Brown2019}   \\
     HR Cam      & 0.09 & 1.1   & 0.41             & 59    & 0.24      & 0.1   & 0.16  & 1.09                    & 0.72              & 6.6                   & 11                                                                                &  \citealt{Maxted1998}   \\
 ESO 330-9 (PN)  & 0.34 & 1.1   & 0.40             & 56    & 0.24      & 0.4   & 0.44  & 0.89                    & 1.7               & 7.0                   & 15                                                                                &  \citealt{Hillwig2017}  \\
    2333+3927    & 0.26 & 1.1   & 0.38             & 43    & 0.24      & 0.3   & 0.36  & 1.18                    & 1.1               & 9.2                   & 17                                                                                &   \citealt{Davis2010}   \\
     FF Aqr      & 1.00 & 1.4   & 0.35             & 25    & 0.32      & 1.4   & 1.3   & 0.16                    & 22                & 22                    & --                                                                                & \citealt{Davis2010}     \\
    V1379 Aql    & 1.00 & 2.3   & 0.30             &  9    & 0.26      & 2.3   & 1.9   & 0.12                    & 43                & 195                   & --                                                                                & \citealt{Davis2010}     \\    
\hline
\end{tabular}
\end{adjustbox}
\end{center}
 \begin{quote}
  \caption{\protect\footnotesize{Observed and derived quantities from several authors ordered by decreasing $M_{\mathrm{c}}$. PN means that the object has a planetary nebula.
  The objects listed by \citet{Schreiber2003}, \citet{DeMarco2008}, \citet{Davis2010} and \citet{Davis2012} were studied previously by several authors and the parameters are vetted compilations. We only list error bars for those values of $M_{\rm c}$ where the error, when available in the original publication, would shift the value across the limit $M_{\mathrm{c}} = 0.47$~\ms\ separating post-RGB ($M_{\mathrm{c}}<0.47$~\ms) from post-AGB ones ($M_{\mathrm{c}} \geq 0.47$~\ms).}} \label{tab:obs_demarco}
 \end{quote}
\end{table*}

\begin{table*}
\begin{center}
\begin{adjustbox}{max width=\textwidth}
\begin{tabular}{+c^c^c^c^c^c^c^c^c^c^c^c^c^c^c}
\hline
Giant type & $q$  & $M_1$ & $M_{\mathrm{c}}$ & $R_1$  & $\lambda$ & $M_2$ & $R_2$ & $R_2/R_{\mathrm{RL,2}}$ & $a_{\mathrm{i}} / R_1$ & $\Omega / \omega^a$ & $a_{\mathrm{f}}^b$ & $|E_{\mathrm{bin}}|$  & $G \left( \frac{M_{\rm c}M_2}{2 a_{\rm f}} - \frac{M_1 M_2}{2 a_{\rm i}} \right)$ & Reference                    \\
           &      & (\ms) & (\ms)            & (\rs)  &           & (\ms) & (\rs) &                         &                        &                     & (\rs)              &
($10^{46}$~erg)       & ($10^{46}$~erg)                                                                   &           \\
\hline                                                                                                                                                                                                                                                                     
AGB        & 0.08 & 5     & 1.0              & 200    & 0.40      & 0.40  & 0.48  & 0.36                    & 1.4                    & 1                   & 4.4                & 
28                    & 16                                                                                & \citealt{Sandquist1998}      \\
AGB        & 0.12 & 5     & 1.0              & 200    & 0.40      & 0.60  & 0.66  & 0.41                    & 1.4                    & 1                   & 4.8                & 
28                    & 22                                                                                & \citealt{Sandquist1998}      \\
AGB        & 0.12 & 5     & 0.94             & 354    & 0.40      & 0.60  & 0.66  & 0.22                    & 1.5                    & 0                   & 8.9                & 
16                    & 11                                                                                & \citealt{Sandquist1998}      \\
AGB        & 0.13 & 3     & 0.7              & 200    & 0.33      & 0.40  & 0.48  & 0.33                    & 1.4                    & 1                   & 4.4                & 
12                    & 11                                                                                & \citealt{Sandquist1998}      \\
AGB        & 0.13 & 3     & 0.7              & 200    & 0.33      & 0.40  & 0.48  & 0.31                    & 1.4                    & 0                   & 4.7                & 
12                    & 10                                                                                & \citealt{Sandquist1998}      \\
RGB        & 0.18 & 4     & 0.7              & 66     & 0.19      & 0.70  & 0.75  & 1.98                    & 1.6                    & 1                   & 1.0                & 
119                   & 88                                                                                & \citealt{Rasio1996}          \\
RGB        & 0.35 & 1     & 0.45             & 243    & 0.20      & 0.35  & 0.43  & 0.06                    & 1.3                    & 0.24                & 21                 & 
1.5                   & 1.2                                                                               & \citealt{Sandquist2000}      \\
RGB        & 0.10 & 1     & 0.45             & 221    & 0.20      & 0.10  & 0.16  & 0.02                    & 1.3                    & 0.14                & 33                 & 
1.7                   & 0.2                                                                               & \citealt{Sandquist2000}      \\
RGB        & 0.18 & 2     & 0.45             & 177    & 0.32      & 0.35  & 0.43  & 0.06                    & 1.3                    & 0.18                & 19                 & 
6.4                   & 1.0                                                                               & \citealt{Sandquist2000}      \\
RGB        & 0.11 & 0.88  & 0.39             & 83     & 0.20      & 0.10  & 0.16  & 0.19                    & 1.0                    & 0                  & 3.1                 & 
3.5                   & 2.2                                                                               & SIM1 (\citealt{Iaconi2018})  \\
RGB        & 0.17 & 0.88  & 0.39             & 83     & 0.20      & 0.15  & 0.22  & 0.21                    & 1.0                    & 0                  & 3.4                 & 
3.5                   & 3.0                                                                               & SIM2 (\citealt{Iaconi2018})  \\
RGB        & 0.34 & 0.88  & 0.39             & 83     & 0.20      & 0.30  & 0.38  & 0.15                    & 1.0                    & 0                  & 7.3                 & 
3.5                   & 2.5                                                                               & SIM3 (\citealt{Iaconi2018})  \\
RGB        & 0.68 & 0.88  & 0.39             & 83     & 0.20      & 0.60  & 0.66  & 0.14                    & 1.0                    & 0                  & 12                  & 
3.5                   & 2.6                                                                               & SIM4 (\citealt{Iaconi2018})  \\
RGB        & 1.02 & 0.88  & 0.39             & 83     & 0.20      & 0.90  & 0.92  & 0.12                    & 1.0                    & 0                  & 17                  & 
3.5                   & 2.1                                                                               & SIM5 (\citealt{Iaconi2018})  \\
RGB        & 0.05 & 1.97  & 0.39             & 66     & 0.32      & 0.10  & 0.16  & 0.27                    & 1.0                    & 0                  & 2.2                 & 
17                    & 2.8                                                                               & SIM6 (\citealt{Iaconi2018})  \\
RGB        & 0.08 & 1.97  & 0.39             & 66     & 0.32      & 0.15  & 0.22  & 0.40                    & 1.0                    & 0                  & 1.8                 & 
17                    & 5.2                                                                               & SIM7 (\citealt{Iaconi2018})  \\
RGB        & 0.15 & 1.97  & 0.39             & 66     & 0.32      & 0.30  & 0.38  & 0.67                    & 1.0                    & 0                  & 1.6                 & 
17                    & 12                                                                                & SIM8 (\citealt{Iaconi2018})  \\
RGB        & 0.30 & 1.97  & 0.39             & 66     & 0.32      & 0.60  & 0.66  & 0.88                    & 1.0                    & 0                  & 1.8                 & 
17                    & 21                                                                                & SIM9 (\citealt{Iaconi2018})  \\
RGB        & 0.46 & 1.97  & 0.39             & 66     & 0.32      & 0.90  & 0.92  & 0.72                    & 1.0                    & 0                  & 2.8                 & 
17                    & 19                                                                                & SIM10 (\citealt{Iaconi2018}) \\
RGB        & 0.30 & 1.97  & 0.39             & 66     & 0.32      & 0.60  & 0.66  & 2.38                    & 1.0                    & 0                  & 0.67                & 
17                    & 62                                                                                & SIM11 (\citealt{Iaconi2018}) \\
RGB        & 0.30 & 1.97  & 0.39             & 66     & 0.32      & 0.60  & 0.66  & 1.83                    & 1.0                    & 0                  & 0.87                & 
17                    & 47                                                                                & SIM12 (\citealt{Iaconi2018}) \\
RGB        & 0.11 & 0.88  & 0.39             & 83     & 0.20      & 0.10  & 0.16  & 0.14                    & 1.0                    & 0                  & 6.1                 & 
3.5                   & 1.6                                                                               & \citealt{Passy2012}          \\
RGB        & 0.17 & 0.88  & 0.39             & 83     & 0.20      & 0.15  & 0.22  & 0.16                    & 1.0                    & 0                  & 7.3                 & 
3.5                   & 2.1                                                                               & \citealt{Passy2012}          \\
RGB        & 0.34 & 0.88  & 0.39             & 83     & 0.20      & 0.30  & 0.38  & 0.12                    & 1.0                    & 0                  & 11                  & 
3.5                   & 1.9                                                                               & \citealt{Passy2012}          \\
RGB        & 0.68 & 0.88  & 0.39             & 83     & 0.20      & 0.60  & 0.66  & 0.10                    & 1.0                    & 0                  & 21                  & 
3.5                   & 1.6                                                                               & \citealt{Passy2012}          \\
RGB        & 1.02 & 0.88  & 0.39             & 83     & 0.20      & 0.90  & 0.92  & 0.09                    & 1.0                    & 0                  & 27                  & 
3.5                   & 1.3                                                                               & \citealt{Passy2012}          \\
RGB        & 0.68 & 0.88  & 0.39             & 100    & 0.20      & 0.60  & 0.66  & 0.08                    & 3.0                    & 0                  & 20                  & 
3.5                   & 1.9                                                                              & \citealt{Iaconi2017}          \\
RGB        & 0.68 & 0.88  & 0.39             & 87     & 0.20      & 0.60  & 0.66  & 0.05                    & 2.5                    & 0                  & 30                  & 
3.5                   & 0.72                                                                              & \citealt{Reichardt2019}      \\ 
RGB        & 0.68 & 0.88  & 0.39             & 91     & 0.20      & 0.60  & 0.66  & 0.06                    & 2.4                    & 0                  & 28                  & 
3.5                   & 0.8                                                                               & \citealt{Reichardt2019}      \\ 
RGB        & 0.68 & 0.88  & 0.39             & 93     & 0.20      & 0.60  & 0.66  & 0.06                    & 2.3                    & 0                  & 28                  & 
3.5                   & 0.77                                                                              & \citealt{Reichardt2019}      \\ 
RGB        & 0.50 & 1.98  & 0.38             & 49     & 0.32      & 0.99  & 0.99  & 0.44                    & 1.0                    & 0.95               & 4.9                 & 
23                    & 7.0                                                                               & \citealt{Ohlmann2016}        \\
RGB        & 0.50 & 1.98  & 0.38             & 52     & 0.32      & 0.99  & 0.99  & 0.67                    & 1.0                    & 0                  & 3.2                 & 
23                    & 15                                                                                & \citealt{Prust2019}          \\
RGB        & 0.50 & 1.98  & 0.38             & 52     & 0.32      & 0.99  & 0.99  & 0.59                    & 1.0                    & 0.95               & 3.6                 & 
23                    & 13                                                                                & \citealt{Prust2019}          \\
RGB        & 0.50 & 1.96  & 0.37             & 48     & 0.32      & 0.98  & 0.98  & 0.26                    & 1.0                    & 0                  & 8.0                 & 
23                    & 1.2                                                                               & \citealt{Chamandy2018}       \\
RGB        & 0.57 & 1.05  & 0.36             & 31     & 0.21      & 0.60  & 0.66  & 0.17                    & 2.0                    & 0.95               & 9.0                 & 
14                    & 2.6                                                                               & \citealt{Ricker2012}         \\
RGB        & 0.27 & 1.18  & 0.36             & 60     & 0.04      & 0.32  & 0.40  & 0.34                    & 2.0                    & 0                  & 3.2                 & 
8.4                   & 6.2                                                                               & \citealt{Nandez2016}         \\
RGB        & 0.31 & 1.18  & 0.36             & 60     & 0.04      & 0.36  & 0.44  & 0.31                    & 2.1                    & 0                  & 3.7                 & 
8.4                   & 6.0                                                                               & \citealt{Nandez2016}         \\
RGB        & 0.34 & 1.18  & 0.36             & 60     & 0.04      & 0.40  & 0.48  & 0.35                    & 2.1                    & 0                  & 3.5                 & 
8.4                   & 7.1                                                                               & \citealt{Nandez2016}         \\
RGB        & 0.23 & 1.38  & 0.36             & 57     & 0.04      & 0.32  & 0.40  & 0.44                    & 2.0                    & 0                  & 2.5                 & 
11                    & 8.0                                                                               & \citealt{Nandez2016}         \\
RGB        & 0.26 & 1.38  & 0.36             & 57     & 0.04      & 0.36  & 0.44  & 0.42                    & 2.0                    & 0                  & 2.8                 & 
11                    & 7.9                                                                               & \citealt{Nandez2016}         \\
RGB        & 0.29 & 1.38  & 0.36             & 57     & 0.04      & 0.40  & 0.48  & 0.41                    & 2.0                    & 0                  & 3.0                 & 
11                    & 8.2                                                                               & \citealt{Nandez2016}         \\
RGB        & 0.20 & 1.59  & 0.36             & 50     & 0.04      & 0.32  & 0.40  & 0.64                    & 1.9                    & 0                  & 1.7                 & 
15                    & 12                                                                                & \citealt{Nandez2016}         \\
RGB        & 0.23 & 1.59  & 0.36             & 50     & 0.04      & 0.36  & 0.44  & 0.65                    & 2.0                    & 0                  & 1.8                 & 
15                    & 13                                                                                & \citealt{Nandez2016}         \\
RGB        & 0.25 & 1.59  & 0.36             & 50     & 0.04      & 0.40  & 0.48  & 0.59                    & 2.0                    & 0                  & 2.1                 & 
15                    & 12                                                                                & \citealt{Nandez2016}         \\
RGB        & 0.18 & 1.8   & 0.36             & 41     & 0.04      & 0.32  & 0.40  & 0.91                    & 1.9                    & 0                  & 1.2                 & 
23                    & 17                                                                                & \citealt{Nandez2016}         \\
RGB        & 0.20 & 1.8   & 0.36             & 41     & 0.04      & 0.36  & 0.44  & 0.90                    & 1.9                    & 0                  & 1.3                 & 
23                    & 17                                                                                & \citealt{Nandez2016}         \\
RGB        & 0.22 & 1.8   & 0.36             & 41     & 0.04      & 0.40  & 0.48  & 0.88                    & 2.0                    & 0                  & 1.4                 & 
23                    & 18                                                                                & \citealt{Nandez2016}         \\     
RGB        & 0.10 & 2     & 0.33             & 44     & 0.32      & 0.20  & 0.40  & 0.39                    & 1.3                    & 0.14               & 2.1                 & 
26                    & 4.7                                                                               & \citealt{Sandquist2000}      \\
RGB        & 0.24 & 1.5   & 0.32             & 26     & 0.29      & 0.36  & 0.44  & 1.25                    & 2.0                    & 0                  & 0.91                & 
30                    & 22                                                                                & \citealt{Nandez2015}         \\
RGB        & 0.27 & 1.2   & 0.32             & 29     & 0.04      & 0.32  & 0.40  & 0.76                    & 2.0                    & 0                  & 1.4                 & 
18                    & 13                                                                                & \citealt{Nandez2016}         \\
RGB        & 0.30 & 1.2   & 0.32             & 29     & 0.04      & 0.36  & 0.44  & 0.76                    & 2.1                    & 0                  & 1.5                 & 
18                    & 13                                                                                & \citealt{Nandez2016}         \\
RGB        & 0.33 & 1.2   & 0.32             & 29     & 0.04      & 0.40  & 0.48  & 0.86                    & 2.1                    & 0                  & 1.4                 & 
18                    & 16                                                                                & \citealt{Nandez2016}         \\
RGB        & 0.23 & 1.4   & 0.32             & 28     & 0.04      & 0.32  & 0.40  & 0.96                    & 2.0                    & 0                  & 1.1                 & 
23                    & 16                                                                                & \citealt{Nandez2016}         \\
RGB        & 0.26 & 1.4   & 0.32             & 28     & 0.04      & 0.36  & 0.44  & 1.03                    & 2.0                    & 0                  & 1.1                 & 
23                    & 18                                                                                & \citealt{Nandez2016}         \\
RGB        & 0.29 & 1.4   & 0.32             & 28     & 0.04      & 0.40  & 0.48  & 1.01                    & 2.0                    & 0                  & 1.2                 & 
23                    & 18                                                                                & \citealt{Nandez2016}         \\
RGB        & 0.20 & 1.6   & 0.32             & 26     & 0.04      & 0.32  & 0.40  & 1.22                    & 1.9                    & 0                  & 0.87                & 
30                    & 20                                                                                & \citealt{Nandez2016}         \\ 
RGB        & 0.23 & 1.6   & 0.32             & 26     & 0.04      & 0.36  & 0.44  & 1.25                    & 2.0                    & 0                  & 0.91                & 
30                    & 22                                                                                & \citealt{Nandez2016}         \\
RGB        & 0.23 & 1.6   & 0.32             & 31     & 0.04      & 0.36  & 0.44  & 1.22                    & 1.6                    & 1                  & 0.93                & 
25                    & 21                                                                                & \citealt{Nandez2016}         \\
RGB        & 0.25 & 1.6   & 0.32             & 26     & 0.04      & 0.40  & 0.48  & 1.26                    & 2.0                    & 0                  & 0.96                & 
30                    & 23                                                                                & \citealt{Nandez2016}         \\
RGB        & 0.18 & 1.8   & 0.32             & 16     & 0.04      & 0.32  & 0.40  & 2.47                    & 1.9                    & 0                  & 0.43                & 
60                    & 42                                                                                & \citealt{Nandez2016}         \\
RGB        & 0.20 & 1.8   & 0.32             & 16     & 0.04      & 0.36  & 0.44  & 2.36                    & 1.9                    & 0                  & 0.48                & 
60                    & 41                                                                                & \citealt{Nandez2016}         \\
RGB        & 0.22 & 1.8   & 0.32             & 16     & 0.04      & 0.40  & 0.48  & 2.28                    & 2.0                    & 0                  & 0.53                & 
60                    & 42                                                                                & \citealt{Nandez2016}         \\
RGB        & 0.35 & 1     & 0.28             & 22     & 0.20      & 0.35  & 0.43  & 0.60                    & 1.3                    & 0.23               & 1.8                 & 
20                    & 8.0                                                                               & \citealt{Sandquist2000}      \\
RGB        & 0.45 & 1     & 0.28             & 18     & 0.20      & 0.45  & 0.53  & 0.52                    & 1.3                    & 0.26               & 2.4                 & 
24                    & 6.3                                                                               & \citealt{Sandquist2000}      \\ 
\hline
\multicolumn{15}{l}{$^a$The variables $\Omega$ and $\omega$ indicate stellar spin and orbital frequencies, so that the column $\Omega / \omega$ indicates the degree of stellar rotation synchronisation with the orbital frequency.} \\
\multicolumn{15}{l}{$^b$The $a_{\mathrm{f}}$ showed are those obtained with the criterion discussed in Section~\ref{sssec:af-ebin}, where available. Where not available they are the values published by the respective authors.} \\
\end{tabular}
\end{adjustbox}
\end{center}
 \begin{quote}
  \caption{\protect\footnotesize{Stellar properties and orbital parameters for the simulations considered in this work, ordered by decresing $M_{\mathrm{c}}$. With respect to Tables~\ref{tab:obs_zorotovic} and \ref{tab:obs_demarco} we added two columns: $a_{\mathrm{i}} / R_1$ and $\Omega / \omega$. These are useful quantities that are readily available only for simulations. The remaining columns are in the same order as those of Tables~\ref{tab:obs_zorotovic} and \ref{tab:obs_demarco}.}} 
  \label{tab:sim}
 \end{quote}
\end{table*}

\label{lastpage}
\end{document}